\documentclass[12pt]{article}
\usepackage{amsbsy,amsmath,amsthm,latexsym,amssymb}
\usepackage{graphicx,stmaryrd,sectsty}
\usepackage{setspace}
\usepackage[export]{adjustbox} 
\usepackage[usenames, dvipsnames]{color} 
\definecolor{NU}{RGB}{82,0,99} 
\newcommand{\bc}{\begin{center}}
\newcommand{\ec}{\end{center}}
\newcommand{\bfr}{\begin{flushright}}
\newcommand{\efr}{\end{flushright}}

\newcommand{\no}{\noindent}
\newcommand{\be}{\begin{enumerate}}
\newcommand{\ee}{\end{enumerate}}
\newcommand{\bi}{\begin{itemize}}
\newcommand{\ei}{\end{itemize}}
\newcommand{\bd}{\begin{description}}
\newcommand{\ed}{\end{description}}
\newcommand{\beq}{\begin{equation}}
\newcommand{\eeq}{\end{equation}}
\newcommand{\bea}{\begin{eqnarray}}
\newcommand{\eea}{\end{eqnarray}}

\newcommand{\bfi}{\begin{figure}}
\newcommand{\efi}{\end{figure}}
\newcommand{\bay}{\begin{array}{l}}
\newcommand{\eay}{\end{array}}

\newcommand{\cref}[1]{(\ref{#1})}   

\sectionfont{\normalsize}
\subsectionfont{\small}
\usepackage{color}
\usepackage{tabularx}
\usepackage{float}
\usepackage[section]{placeins}

\begin{document}

\begin{titlepage}
\clearpage\thispagestyle{empty}
\noindent
\hrulefill
\begin{figure}[h!]
\centering
\includegraphics[width=2 in]{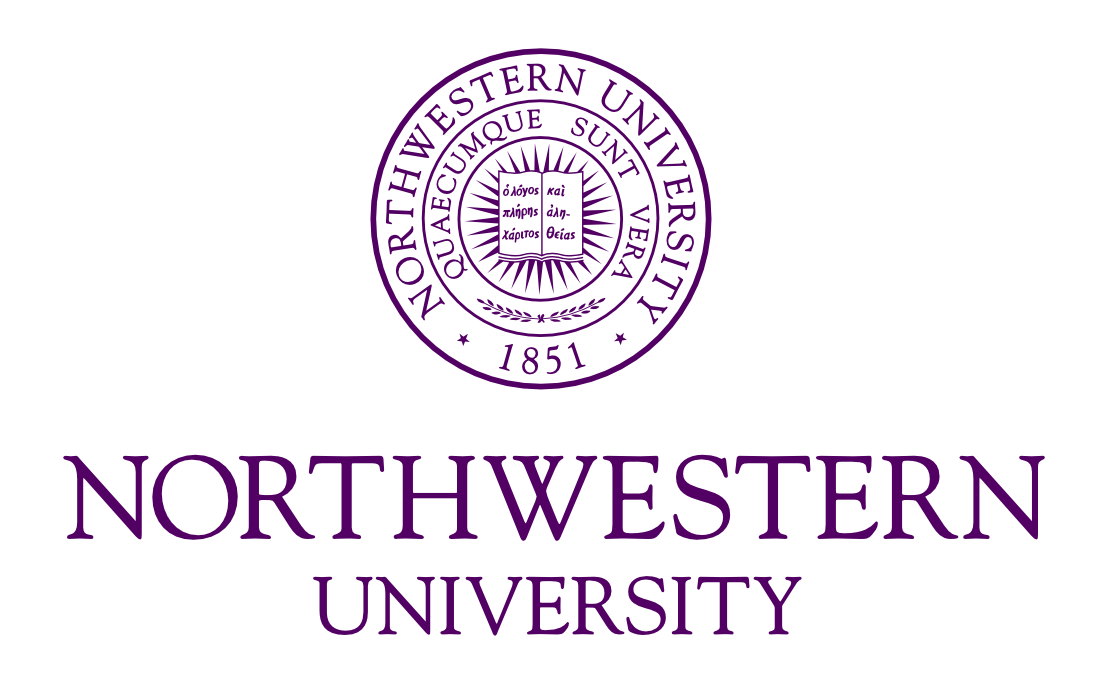}
\end{figure}
\begin{center}
{\color{NU}{
{\bf Center for Sustainable Engineering of Geological and Infrastructure Materials (SEGIM)} \\ [0.1in]
Department of Civil and Environmental Engineering \\ [0.1in]
McCormick School of Engineering and Applied Science \\ [0.1in]
Evanston, Illinois 60208, USA
}
}
\end{center} 
\hrulefill \\ \vskip 2mm
\vskip 0.5in
\begin{center}
{\large {\bf A Novel Material for In Situ Construction on Mars: \\Experiments and Numerical Simulations}}\\[0.5in]
{\large {\sc Lin Wan, Roman Wendner, Gianluca Cusatis}}\\[0.75in]
{\sf \bf SEGIM INTERNAL REPORT No. 15-12/487A}\\[0.75in]
\end{center}
\noindent {\footnotesize {{\em Accepted for publication in Construction and Building Materials \hfill May 2016} }}
\end{titlepage}

\newpage
\clearpage \pagestyle{plain} \setcounter{page}{1}

\begin{center}
{\large {\bf A Novel Material for In Situ Construction on Mars: \\Experiments and Numerical Simulations}}
\\[9mm]
{\bf  {\bf  By \\
Lin Wan \footnote{PhD, Researcher, Department of Civil and Environmental Engineering, Northwestern University, 2145 Sheridan Rd. Evanston IL, 60208 USA. E-mail: lin.wan@u.northwestern.edu},
Roman Wendner \footnote{Director Christian Doppler Laboratory LiCRoFast, Department of Civil Engineering and Natural Hazards, University of Natural Resources and Life Sciences (BOKU) Vienna. E-mail: roman.wendner@boku.ac.at},
Gianluca Cusatis$^*$ \footnote{Corresponding Author: Associate Professor, Department of Civil and Environmental Engineering, Northwestern University, 2145 Sheridan Rd. Evanston IL, 60208 USA. E-mail: g-cusatis@northwestern.edu, Phone: (847)-491-4027} 
}  }
\vspace{15mm}

A Paper Accepted for Publication in

Construction and Building Materials

May 4, 2016

Corresponding Author

Gianluca Cusatis

Associate Professor

Department of Civil and Environmental Engineering

Tech Building A125 

Northwestern University

2145 Sheridan Rd 

Evanston, IL 60208

Tel: (847) 491-4027

E-mail: g-cusatis@northwestern.edu

Website: http://www.cusatis.us

\end{center}

\newpage

\no {\bf   Abstract:}\\ 

A significant step in space exploration during the 21st century will be human settlement on Mars. Instead of transporting all the construction materials from Earth to the red planet with incredibly high cost, using Martian soil to construct a site on Mars is a superior choice. Knowing that Mars has long been considered a \textquotedblleft sulfur-rich planet\textquotedblright, a new construction material composed of simulated Martian soil and molten sulfur is developed. In addition to the raw material availability for producing sulfur concrete and a strength reaching similar or higher levels of conventional cementitious concrete, fast curing, low temperature sustainability, acid and salt environment resistance, 100\% recyclability are appealing superior characteristics of the developed Martian Concrete. In this study, different percentages of sulfur are investigated to obtain the optimal mixing proportions. Three point bending, unconfined compression and splitting tests were conducted to determine strength development, strength variability, and failure mechanisms. The test results show that the strength of Martian Concrete doubles that of sulfur concrete utilizing regular sand. It is also shown that the particle size distribution plays an important role in the mixture's final strength. Furthermore, since Martian soil is metal rich, sulfates and, potentially, polysulfates are also formed during high temperature mixing, which might contribute to the high strength. The optimal mix developed as Martian Concrete has an unconfined compressive strength of above 50~MPa. The formulated Martian Concrete is simulated by the Lattice Discrete Particle Model (LDPM), which exhibits excellent ability in modeling the material response under various loading conditions.

\section{Introduction}\no


Sulfur has been used as a molten bonding agent for quite a long time in human history. The use of sulfur was mentioned in literature of ancient India, Greece, China and Egypt \cite{Sheppard1975}. For example, sulfur was one of the raw materials to manufacture gunpowder by ancient Chinese \cite{Buchanan2006}; sulfur was also used to anchor metal in stone during the 17th century \cite{Rybczynski1974}. Starting in the 1920s, sulfur concrete has been reported to be utilized as a construction material \cite{ACI1998}. Various researchers and engineers studied and succeeded in obtaining high-strength and acid-resistant sulfur concretes \cite{Bacon1921, Kobbe1924, Duecker1934}. In the late 1960s, Dale and Ludwig pointed out the significance of well-graded aggregate in obtaining optimum strength \cite{Dale1966, Dale1968}. 

When elemental sulfur and aggregate are hot-mixed, cast, and cooled to prepare sulfur concrete products, the sulfur binder, on cooling from the liquid state, first crystallizes as monoclinic sulfur (S$_\beta$) at 238~$^\circ$F (114~$^\circ$C). On further cooling to below 204~$^\circ$F (96~$^\circ$C), S$_\beta$ starts to transform to orthorhombic sulfur (S$_\alpha$), which is the stable form of sulfur at ambient room temperatures \cite{1976}. This transformation is rapid, generally occurring in less than 24 hours and resulting in a solid construction material. However, since S$_\alpha$ is much denser than S$_\beta$, high stress and cavities can be induced by sulfur shrinkage. Hence, durability of unmodified sulfur concrete is a problem when exposed to humid environment or after immersion in water. In the 1970s, researchers developed techniques to modify the sulfur by reacting it with olefinic hydrocarbon polymers \cite{Vroom1977, Vroom1981}, dicyclopentadiene (DCPD) \cite{Leutner1977, Bordoloi1978, Gregor1978, Schneider1981, McBee1982}, or other additives and stabilizers \cite{Bright1978, Gillott1980, Woo1983} to improve durability of the product. Since then, commercial production and installation of corrosion-resistant sulfur concrete has been increasing, either precast or installed directly in industrial plants where portland cement concrete materials fail from acid and salt corrosion \cite{ACI1998}. 


For earth applications, well developed sulfur concrete features (1) improved mechanical performance: high compressive $\&$ flexural strength, high durability, acid $\&$ salt water resistant, excellent surface finish and pigmentation, superior freeze/thaw performance; (2) cost benefits: faster setting-solid within hours instead of weeks, increased tolerance to aggregate choice; and (3) environmentally friendly profile: reduced CO$_2$ footprint, no water requirements, easily obtainable sulfur as a byproduct of gasoline production, recyclable via re-casting, compatible with ecosystem, e.g. for marine applications. Current pre-cast sulfur concrete products include, but are not limited to, flagstones, umbrella stand, counterweights for high voltage lines, and drainage channel \cite{Britton2010}. 

For example, in January 2009, around 80 meters sewage pipeline in the United Arab Emirates was removed and replaced by sulfur concrete. In the same time period, a total of 215 fish reef blocks made of sulfur concrete (2.2 tons/block) were stacked at a depth of 15 meters, 6 kilometers off the coast of UAE \cite{Lida2009}. With regular concrete fish reefs, the growth of algae and shells takes time because concrete is alkaline. However, since sulfur concrete is practically neutral in alkalinity, algae and shell growth was observed soon after installation. 


While sulfur concrete found its way into practice as an infrastructure material, it is also a superior choice for space construction considering the very low water availability on the nearby planets and satellites \cite{Casanova1997}. After mankind stepped on the lunar surface in 1969, space agencies have been planning to go back and build a research center on the moon. Since local material is preferred to reduce expenses, starting in the early 1990s, NASA and collaborative researchers studied and developed lunar concrete using molten sulfur. Around the year 1993, Omar \cite{Omar1993} made lunar concrete by mixing lunar soil simulant with different sulfur ratio ranging from 25\% up to 70\% and found the optimum mix with 35\% sulfur to reach a compressive strength of 34~MPa. Later he added 2\% of steel fibers to the mixes and increased the optimum strength to 43~MPa. However, lunar concrete has serious sublimation issues because of the near-vacuum environment on the moon. In 2008, Grugel and Toutanji \cite{NASA2008, Grugel2008, Grugel2012} reported experimental results of two lunar concrete mixes: (1) 35\% sulfur with 65\% lunar soil simulant JSC-1, and (2) 25\% sulfur and 20\% silica binder mixture with 55\% JSC-1. The two mixtures, similar in strength ($\sim$ 35~MPa), revealed a continuous weight loss due to the sublimation of sulfur when placed in a vacuum environment, 5$\times$10$^{-7}$ torr, at 20~$^\circ$C for 60 days. Based on the measurements, it was predicted that sublimation of a 1 cm deep layer from the two sulfur concrete mixes would take 4.4 and 6.5 years respectively. The sublimation rate varied from rapid at the high lunar temperatures (\textless 120~$^\circ$C) to essentially nonexistent at the low lunar temperatures (-180~$^\circ$C to -220~$^\circ$C). However, the low temperature on the moon is too harsh to maintain intact the mechanical properties of sulfur concrete. After cycled 80 times between -191~$^\circ$C (-312~$^\circ$F) and 20~$^\circ$C (68~$^\circ$F), the samples failed at about 7~MPa under compression, which is about 1/5 of the average strength, 35~MPa, of the non-cycled samples. 

While the moon is the closest and only satellite of earth, its near-vacuum environment, broad temperature range and long day-night rhythm, about 30 earth days, are not the most adequate for human settlement. Venus is the closest planet to Earth, however it is also the hottest planet in the solar system with an average surface temperature over 400~$^\circ$C \cite{Shalygin2015}, making it uninhabitable for humans. Mars, on the other hand, is not too hot or too cold, and has an atmosphere to protect humans from radiation. Its day/night rhythm is very similar to that on Earth: a Mars day is about 24 hours and 37 minutes \cite{Lodders1998}. Thus, Mars is the most habitable planet in the solar system after Earth. In recent years, many countries, including the U.S., China, and Russia, announced to launch manned Mars missions in the next decades. Due to the dry environment on Mars, sulfur concrete concept is a superior choice for building a human village on the red planet. Studies of Martian meteorites suggest elevated sulfur concentrations in the interior, and Martian surface deposits contain high levels of sulfur (SO$_3$ up to ~37 wt\%, average ~6 wt\%), likely in the forms of sulfide minerals and sulfate salts \cite{King2010}. Except of the easiest option of finding a sulfur mine on Mars, like the one in Sicily on Earth, elemental sulfur can be extracted from sulfides or sulfates through various chemical and physical processes, for example, by heating up the sulfur compounds \cite{Zhang1986}. NASA has advanced programs on In Situ Resources Utilization (ISRU) \cite{AIAA2007} for this specific purpose. Moreover, the atmospheric pressure (0.636 kPa) \cite{Barlow2008} as well as temperature range ($\leq$ 35~$^\circ$C) are highly suitable for the application of sulfur concrete. As shown in Fig.~\ref{sulfur} \cite{NASA2008}, the most possible construction site on Mars has environmental conditions in the Rhombic (stable) state of sulfur and is three orders of magnitude in pressure above the solid-vapor interface. Thus, sublimation is not an issue and a relatively warm area can be selected as the construction site. Furthermore, with the temperature on Mars lower than 35~$^\circ$C, the drawback of sulfur concrete melting at high temperature will not be an issue for initial constructions such as shelters and roads while certainly might be of concern for long term settlements in which fire resistance would be important.

To let the thoughts become facts, a new construction material using simulated Martian soil and molten sulfur is developed in this study. Different percentages of sulfur are studied to obtain the optimal mixing proportions. Through mechanical tests, it is found that Martian Concrete have much higher strengths than sulfur concrete utilizing regular sand. Sieve analysis and chemical analysis provide possible explaination for the higher strength of Martian Concrete: the Martian soil simulant has a better particle size distribution, it is also rich in metal elements, which react with sulfur, forming polysulfates and possibly enhancing strengths. Mechanical simulations of Martian Concrete are then carried out using the state-of-art Lattice Discrete Particle Model with excellent simulation of Martian Concrete mechanical properties.

\bfi[htbp]
   \centering
   \includegraphics[width=5in]{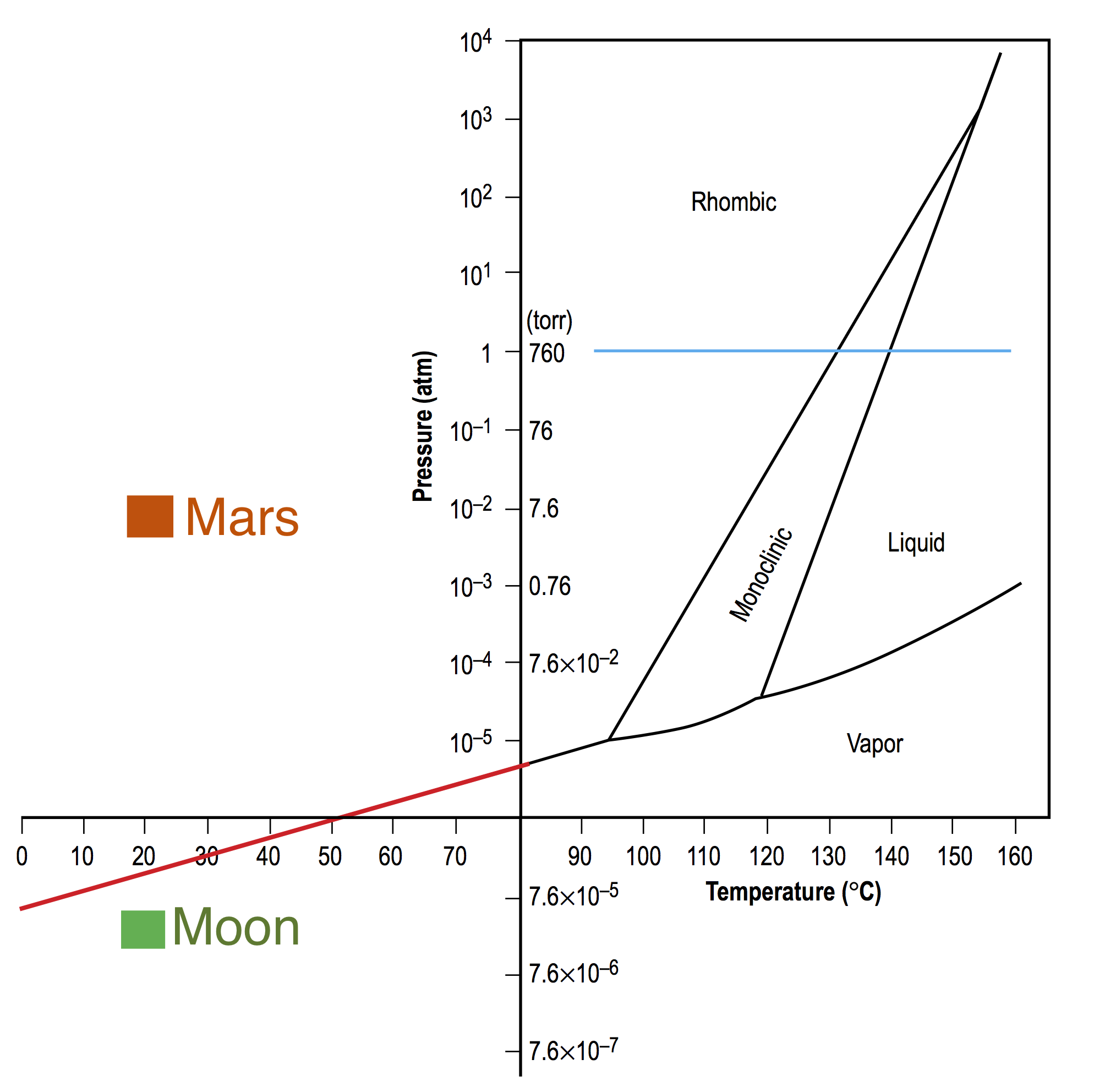} 
   \caption{Sulfur phase diagram with labeled environmental conditions on Mars and Moon \cite{NASA2008}}
   \label{sulfur}
\efi

\section{Experimental Study of Martian Concrete}\no


Sulfur concrete products are manufactured by hot-mixing sulfur and aggregate. The sulfur binder first crystalizes as monoclinic sulfur (S$_\beta$), and then the mixture cools down while sulfur transforms to the stable orthorhombic polymorph (S$_\alpha$), achieving a reliable construction material. While sulfur is commercially available, Martian soil simulant JSC Mars-1A \cite{Mars1A} was obtained in replacement of Martian soil to develop a feasible Martian Concrete. Table 1 lists the major element composition of the simulant. As seen, the Martian soil simulant, resembling the actual Martian soil \cite{Mars1}, is rich with metal element oxides, especially aluminium oxide and ferric oxide. In this study, various percentages of sulfur are mixed with JSC Mars-1A in a heated mixer at above 120~$^\circ$C. Temperature measurements are performed during mixing to ensure sulfur melting. Then the mixture is transferred to 25.4$\times$25.4$\times$127~mm (1$\times$1$\times$5~in) aluminum formwork when it reached flowable state or best mixing conditions. Afterwards the material was let to cool down at room temperature, about 20~$^\circ$C. Martian soil simulant Mars-1A of maximum 5~mm aggregate size was first used for casting, however the specimens showed many voids and uneven surfaces due to the large aggregate, see Fig.~\ref{msize}a. Sulfur cannot be ensured to fill the large number of big voids or to surround and bind all large aggregates, especially on the specimen surface. Afterwards, only Mars-1A of maximum 1~mm aggregate size was utilized to achieve Martian Concrete (MC) with flat and smooth surfaces, see Fig.~\ref{msize}b. Mechanical tests were conducted after 24 hours, and these included unconfined compression, notched and unnotched three-point-bending (TPB), and splitting (Brazilian) tests. Beams of dimensions 25.4$\times$25.4$\times$127~mm (1$\times$1$\times$5~in) are used for TPB tests, which are then cut to 25.4~mm (1~in) cubes for compression and splitting tests.

\begin{table}[ht]
\centering
\begin{tabular}{l c}
\multicolumn{2}{l}{Table 1: Major Element Composition of Martian Regolith Simulant JSC Mars-1A \cite{Mars1A}}\\
\hline
Major Element Composition & \% by Wt.\\
\hline
Silicon Dioxide (SiO$_2$) & 34.5-44\\
Titanium Dioxide (TiO$_2$) & 3-4\\
Aluminum Oxide (Al$_2$O$_3$) & 18.5-23.5\\
Ferric Oxide (Fe$_2$O$_3$) & 9-12\\
Iron Oxide (FeO) & 2.5-3.5\\
Magnesium Oxide (MgO) & 2.5-3.5\\
Calcium Oxide (CaO) & 5-6\\
Sodium Oxide (Na2O) & 2-2.5\\
Potassium Oxide (K$_2$O) & 0.5-0.6\\
Manganese Oxide (MnO) & 0.2-0.3\\
Diphosphorus Pentoxide (P$_2$O$_5$) & 0.7-0.9\\
\hline
\end{tabular}
\end{table}

\bfi
\centering
(a) \includegraphics[height=1in,valign=t]{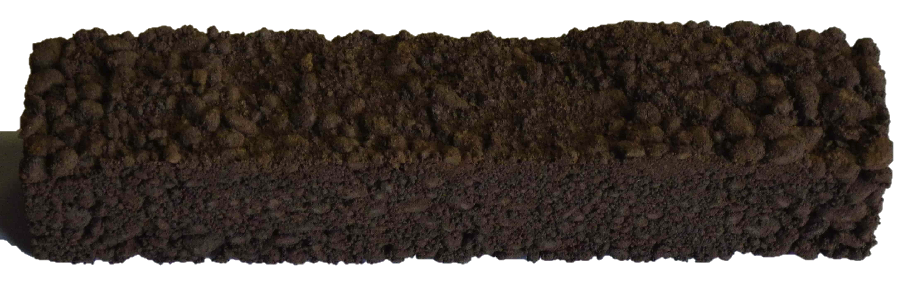}
(b) \includegraphics[height=1in,valign=t]{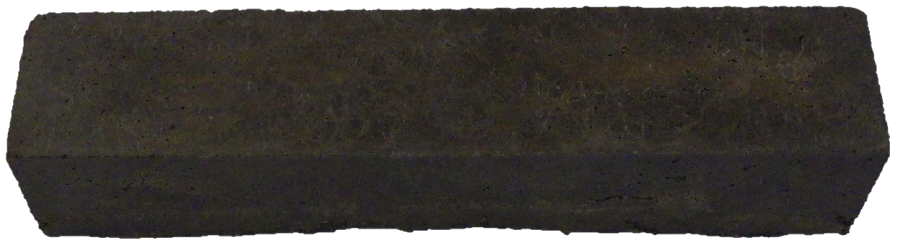}
\caption{Martian Concrete beams utilizing Martian soil simulant with (a) maximum 5~mm aggregate,  and (b) maximum 1mm aggregate}
\label{msize}
\efi

\subsection{Unconfined Compression Test}

Unconfined compression tests were performed in a closed loop servo-hydraulic load frame with a maximum capacity of 489 kN (110 kips). Stroke/displacement control with a loading rate of 0.003~mm/s was applied. In order to ensure consistent and accurate test results, a Standard Operation Procedure (SOP) for testing was created. The test protocol was first filled with the relevant details, which include Vernier Caliper measurements of each dimension (average of 2 $\sim$ 4 measurements), the initial weight, the label of the specimen, control mode, loading rate, and start time of loading. Pictures were taken to document the initial condition of the specimen, during test and post test states. A preload of approximately 1-5 $\%$ of the expected peak-load was applied before the actual test commenced. 

Specimens used for unconfined compression tests were 25.4~mm (1~in) cubes cut from the undamaged parts of 25.4$\times$25.4$\times$127~mm (1$\times$1$\times$5~in) beams, see Fig.~\ref{cube}a. The cubes were cut out of the 62~mm (2.5~in) long failed half's at the center between bending test support point and fracture surface. Typical cone type of failure is observed of Martian Concrete under unconfined compression, as shown in Fig.~\ref{cube}b. 

The studied sulfur ratio for Martian Concrete under compression tests ranged from 35 wt\% to 60 wt\%. Compressive strength versus percentage of sulfur is shown in Fig.~\ref{percp} (circles), revealing an optimum percentage around 50$\%$ ($\pm$ 2.5\%). Furthermore, the test results indicate that recast can further increase strength of the material. For 50$\%$ sulfur batches, recast made compressive strength go up from 48~MPa to about 58~$\sim$~63~MPa, which is roughly a 20 $\sim$ 30$\%$ increase, see Fig.~\ref{percp} labeled as \textquotedblleft Mars1A~1mm~R.\textquotedblright{}. Furthermore, better mixing and applying pressure while placing the material in formwork facilitates material strength. In the experimental campaign of this study, a well distributed pressure was manually added to the mixture in formwork, and thus the pressure was not quantified. Making the mixture compact facilitates formation of sulfur bonds and also reduces the number and size of cavities of the final product. Average compressive stress-strain curves for MC with sulfur ratio ranging from 40\% to 60\% are plotted in Fig.~\ref{alle}a. Stress is calculated as $P$/$A$, where $P$ is load and $A$ is the area of the cross section; strain is calculated as $\Delta h/h$, where $h$ is the height of the specimen. The stress-strain curves feature a typical almost-linear behavior up to the peak and a long stable softening post-peak.

While Martian Concrete has a high strength of over 50~MPa with relatively high percentage of sulfur, sulfur concrete made of regular sand (Sand Concrete, SC) was cast and tested as well for comparison. With the same dimension of 25.4~mm (1~in), SC cubes were cast with a sulfur ratio in the range of 15\% $\sim$ 35\%. Sand with a maximum aggregate size of 11~mm was first utilized. Then for comparison purposes, maximum 1~mm sand, sieved from the coarser sand, was used as well. Following the same test procedure, SC specimens were tested under unconfined compression loads. As shown in Fig. \ref{percp}, the best percentage of sulfur for SC was found to be about 25\% for both fine (crosses in Fig.~\ref{percp}) and coarse (squares in Fig.~\ref{percp}) mixes, having 24.5~MPa and 28.3~MPa compressive strength, respectively. The results obtained on the SC mixes are consistent with the existing literature on standard sulfur concrete \cite{ACI1998}. When the aggregate size distribution of the fine sand was modified based upon the particle size distribution of Mars-1A simulant, its SC mix's stregnth had a 29\% jump to 31.5~MPa, see Fig.~\ref{percp} labeled as \textquotedblleft Sand1A 1mm\textquotedblright{} and marked with a diamond symbol. This indicates and confirms the significance of the particle size distribution in order to obtain an optimum material strength.

\bfi
\centering
(a) \includegraphics[height=1.9in,valign=t]{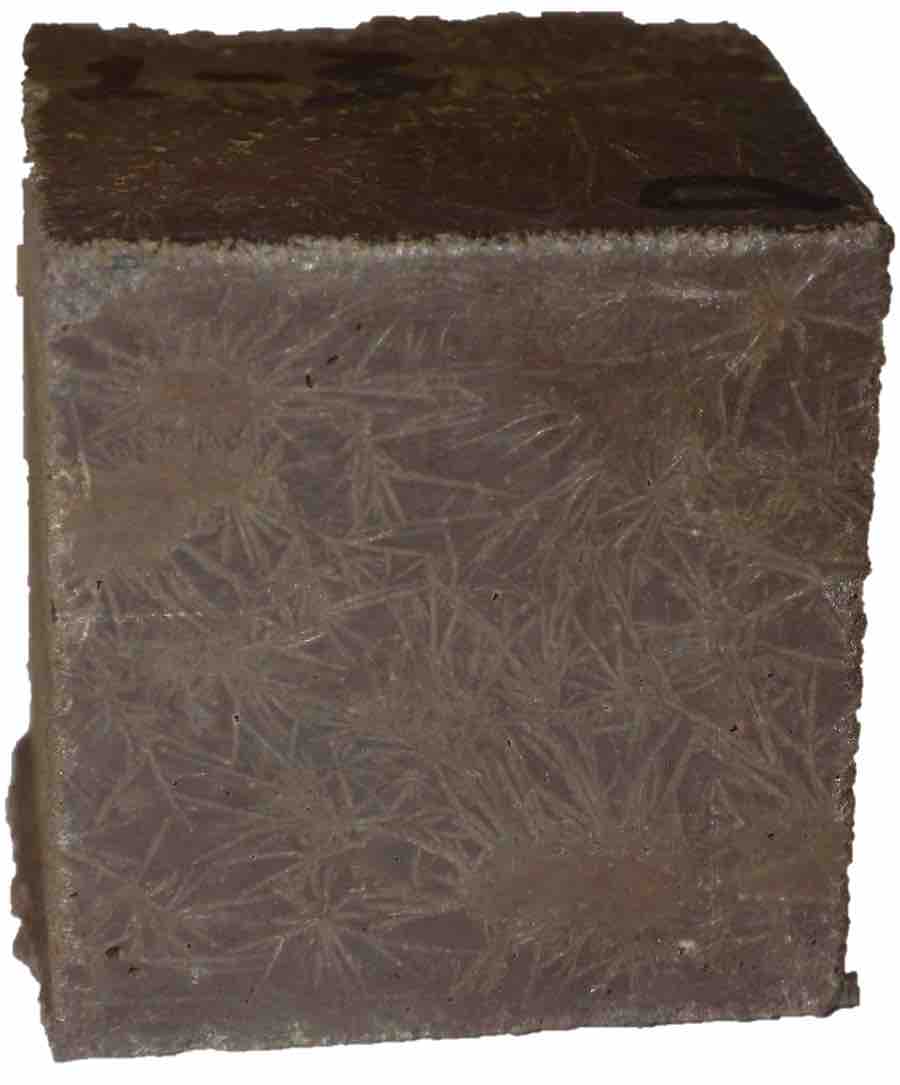}
(b) \includegraphics[height=1.5in,valign=t]{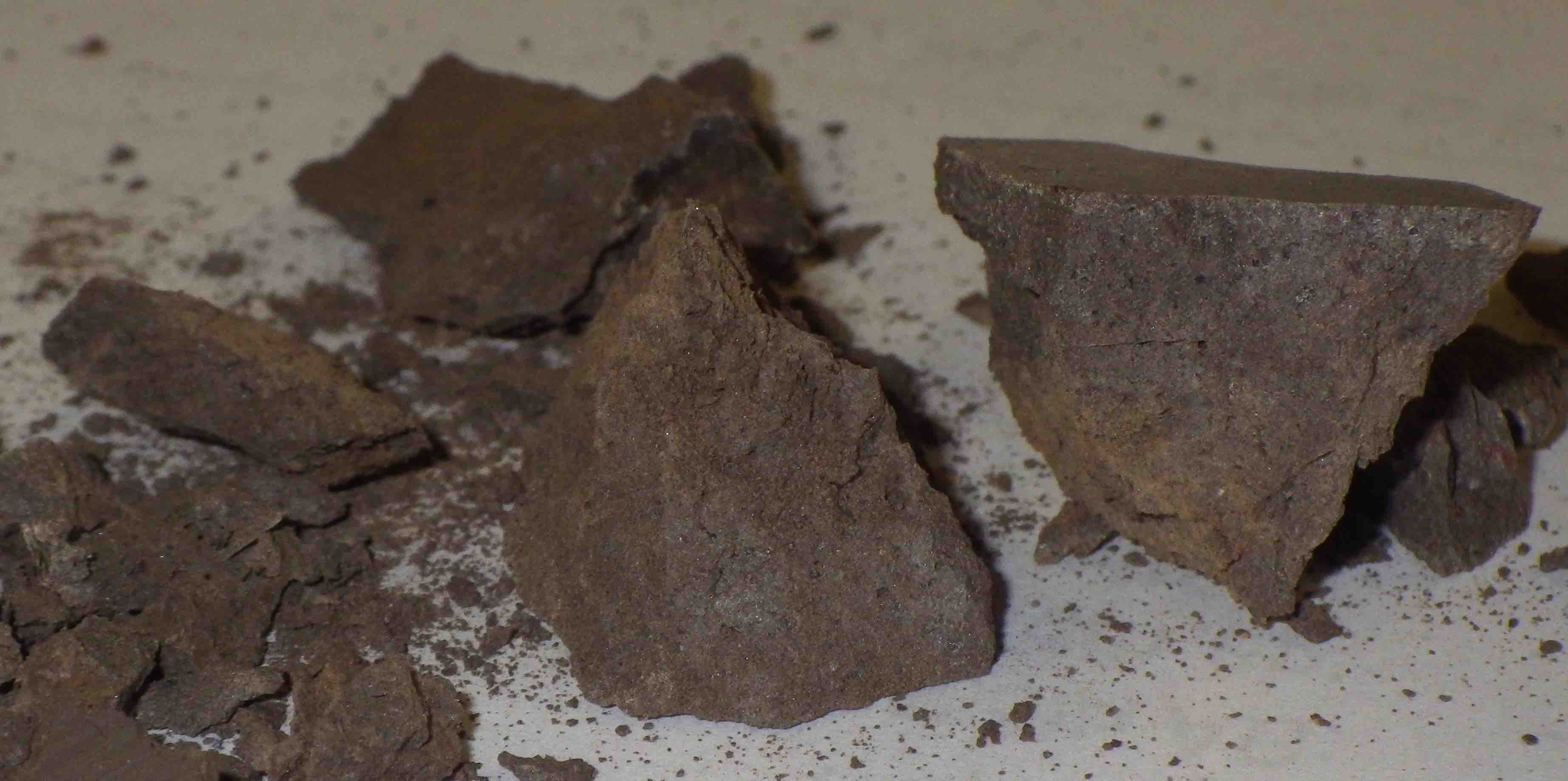}
\caption{Cube specimen (a) before and (b) after unconfined compression test}
\label{cube}
\efi

\bfi
\centering
\includegraphics[width=6in,valign=t]{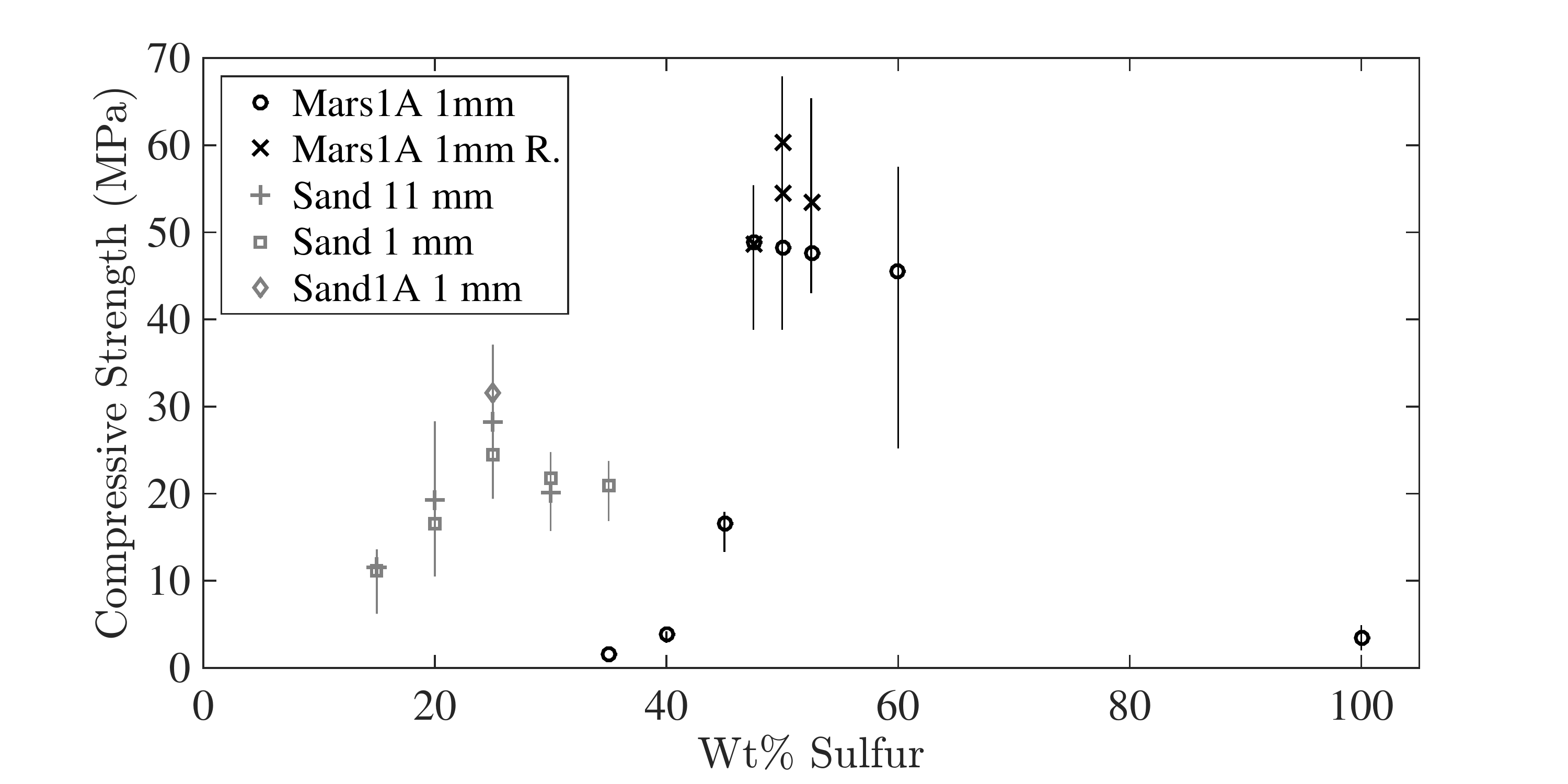}
\caption{Compression strength variation as a function of percentage of sulfur for Martian Concrete}
\label{percp}
\efi

\bfi
\centering
(a) \includegraphics[width=3.25in,valign=t]{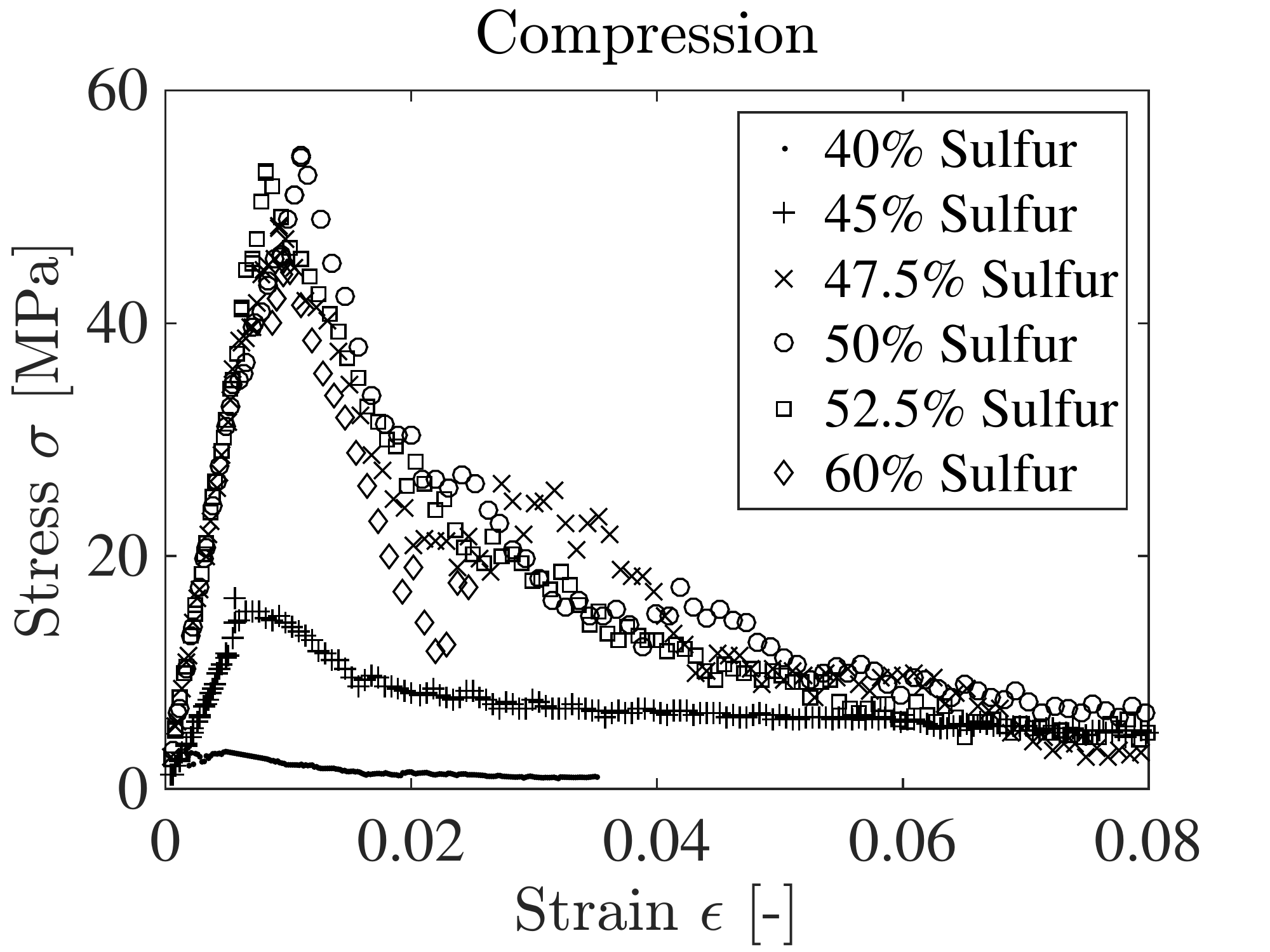}
(b) \includegraphics[width=3.25in,valign=t]{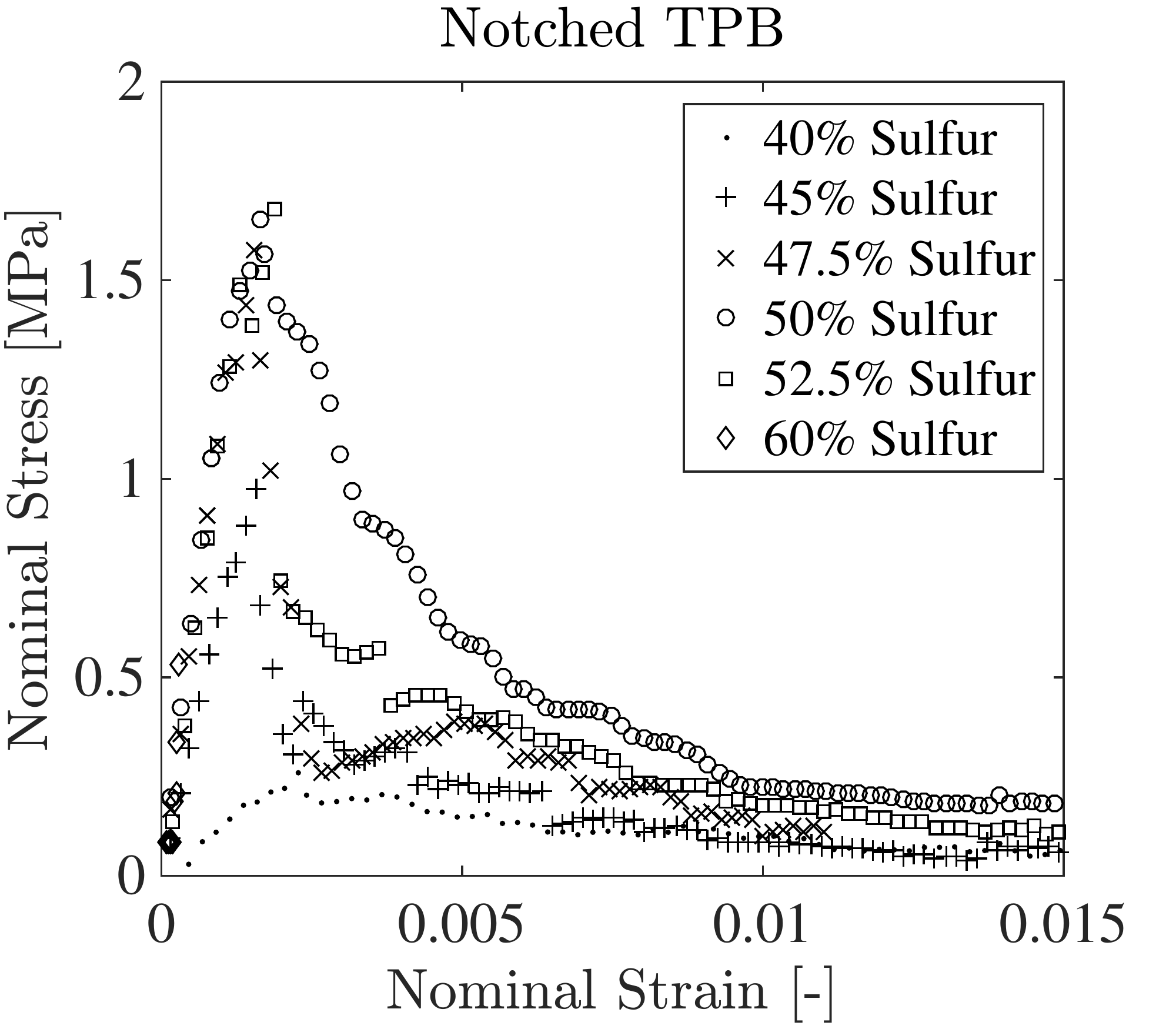}
\caption{Comparison of the response for Martian Concrete with various sulfur ratio by (a) compression and (b) 50\% notched three point bending tests}
\label{alle}
\efi

\subsection{Particle Size Distribution Analysis}

While 25\% of elemental sulfur works the best for both mixes with regular sand, they also both have much lower strength compare to Martian Concrete. To study the influence of aggregates and the corresponding particle size distribution (PSD) on material strength, sieve analyses of Mars-1A (maximum 1~mm aggregate size) as well as regular sand (maximum 11~mm aggregate size) were conducted. Also included in the PSD analysis were the recommended PSDs by ASTM and AASHTO standards for mixing sulfur concrete \cite{ACI1998}. In Fig.~\ref{sieve}, the normalized distributions of Mars-1A, regular sand, the ASTM D 3515 and AASHTO recommended PSD ranges as well as Fuller's law with power 1/2 are plotted and compared. Overall, the PSD of Mars-1A falls well in the recommended PSD range according to standards and is relatively close to Fuller's law, while the PSD of regular sand misses the recommended PSD range and also deviates from Fuller's law. While this finding explains partly the difference in the measured strength of MC and SC, it cannot justify the more than doubled strength of MC compared to SC.

\bfi
   \centering
   \includegraphics[width=6in]{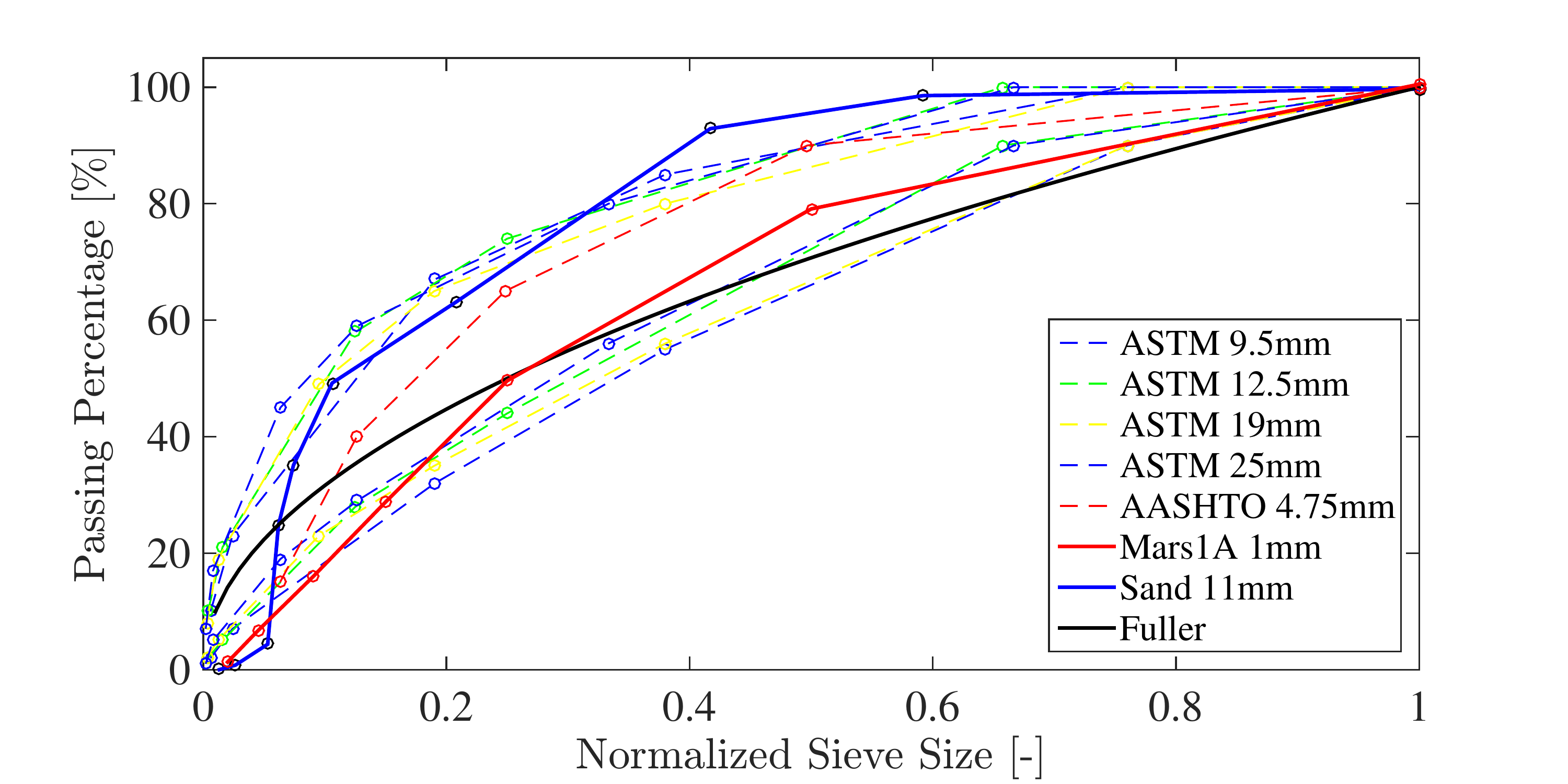} 
   \caption{Particle size distribution (PSD) study of Martian soil simulant and regular sand as well as ASTM and AASHTO recommended PSD for mixing sulfur concrete}
   \label{sieve}
\efi

\subsection{Microscopy Study}

In addition to the PSD of aggregate, other factors must play a role concerning the final strength obtained in MC experiments. Fig.~\ref{m1}~\&~\ref{m2} show the microscope study of Martian Concrete (MC) and sulfur concrete with regular sand (SC) with optimal compositions. By comparing the particles of MC and SC in the mesostructure pictures, a few observations are in order. Firstly, the visible average particle size of MC is much smaller than that of SC after hot mixing, although both mixes use aggregate with maximum particle size up to 1~mm. After casting and curing, the aggregate particles and their sizes can be well distinguished for SC; on the contrary, the majority of MC particles are below 500 microns. Secondly, the MC mix has many red areas, dark spots and almost no voids, while the SC mix shows distinguishably yellow areas of sulfur,  opaque orange to dark red spots related to sand particles and a number of voids of around 200 microns. These observations, along with preliminary X-ray photoelectron spectroscopy (XPS) tests, suggest that the metal elements in Mars-1A react with sulfur during hot mixing, forming sulfates and polysulfates, and altering the PSD of aggregates to lower ends, which further enhance the MC strength. SC does not have such phenomena because silica sand does not react with sulfur at the aforementioned casting conditions. In other words, in MC aggregate is chemically active whereas in SC is inert and sulfur only serves as \textquotedblleft glue\textquotedblright{} for the sand particles. The existence of sulfates and polysulfates in MC are qualitatively confirmed by XPS by analyzing the chemical state of sulfur and individual metal elements within 900~micron-diameter areas of a thin MC sample. Definitely, further research is needed to clearly identify the chemical products characterizing MC internal structure.


\bfi
\centering
(a) \includegraphics[width=3in,valign=t]{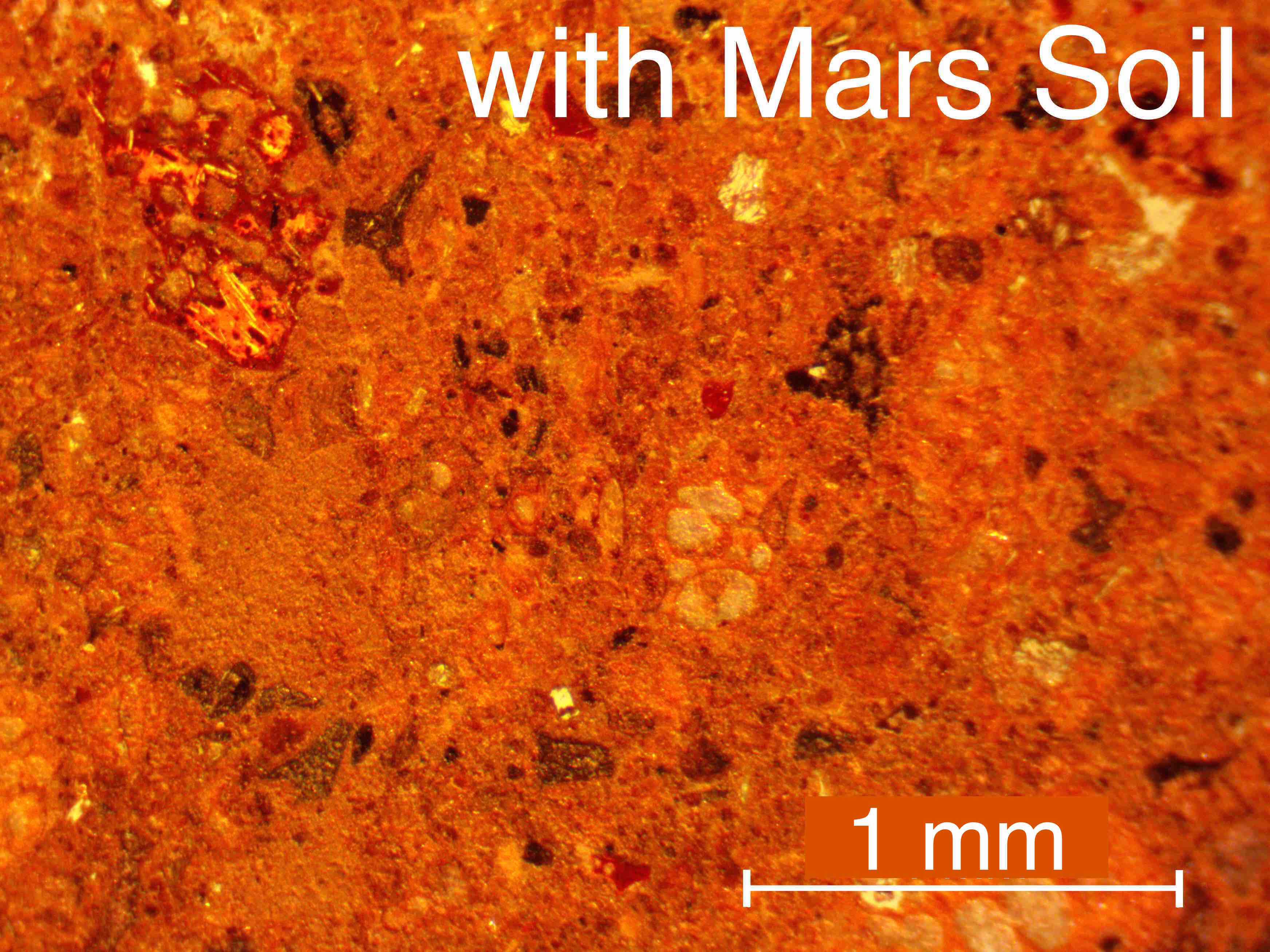}
(b) \includegraphics[width=3in,valign=t]{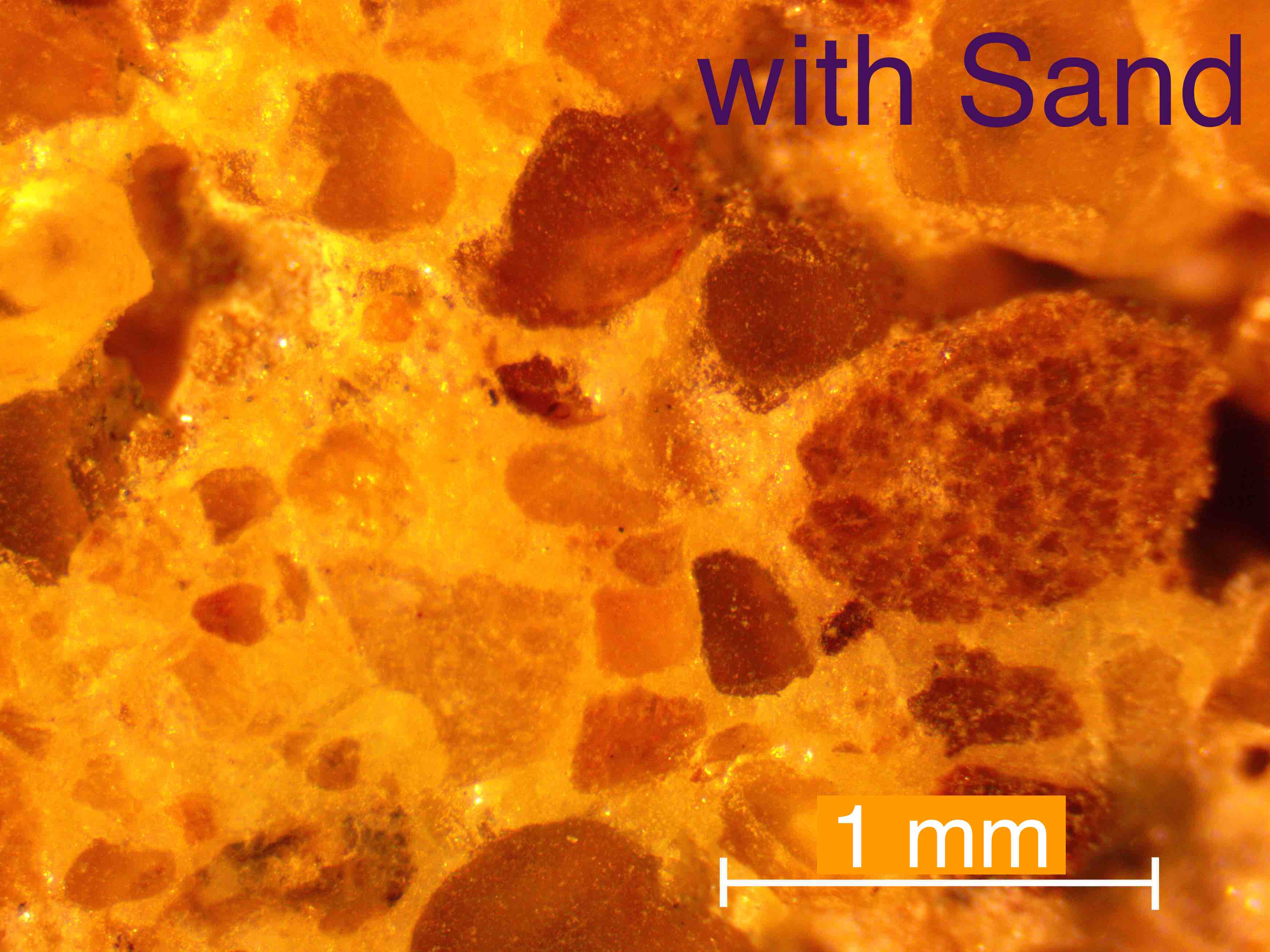}
\caption{Microscopy study of sulfur concrete on 1~mm scale with compositions of (a) 50\% sulfur and 50\% Martian soil simulant (b) 25\% sulfur and 75\% regular sand and a maximum particle size of 1~mm}
\label{m1}
\efi

\bfi
\centering
(a) \includegraphics[width=3in,valign=t]{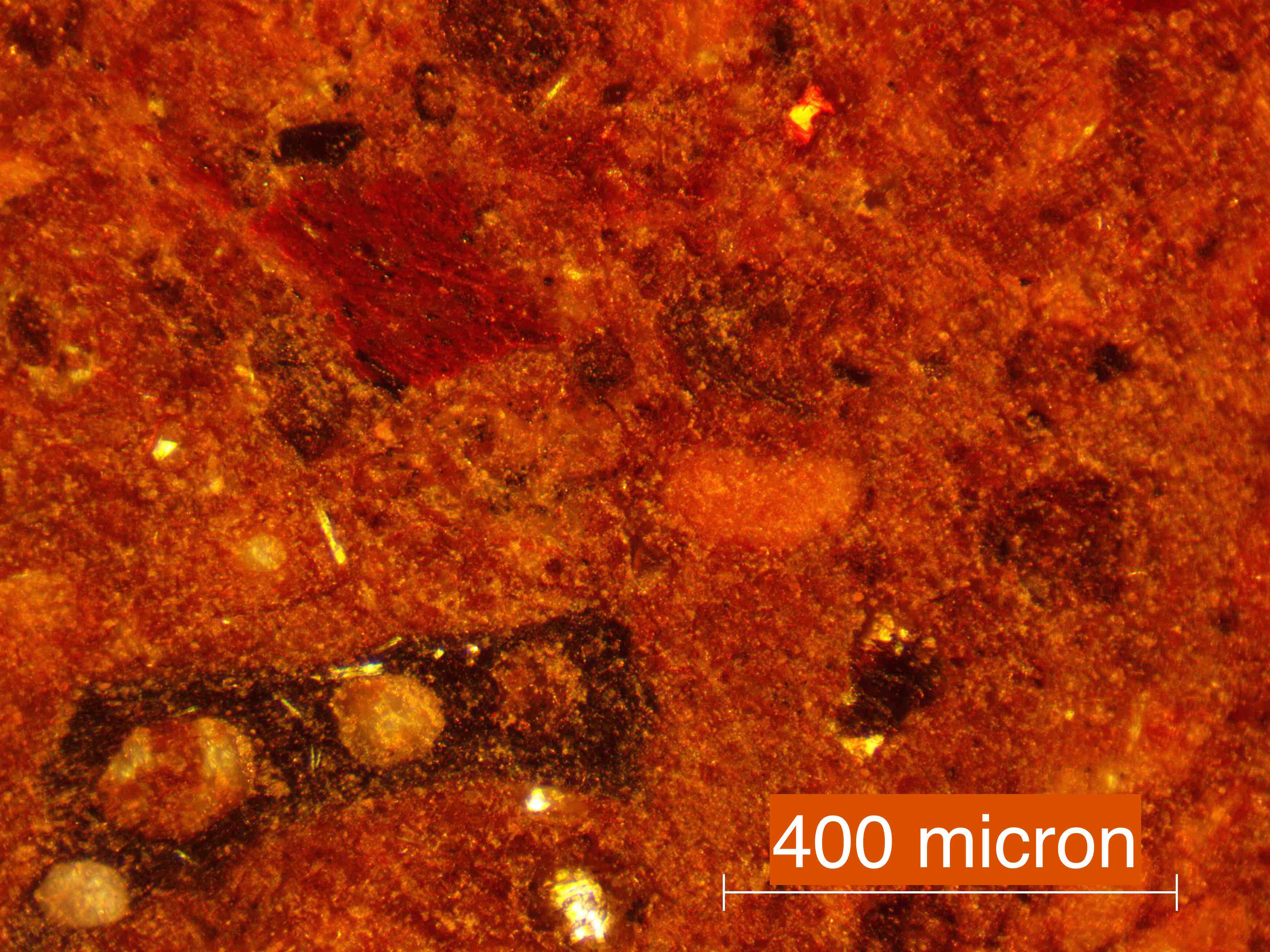}
(b) \includegraphics[width=3in,valign=t]{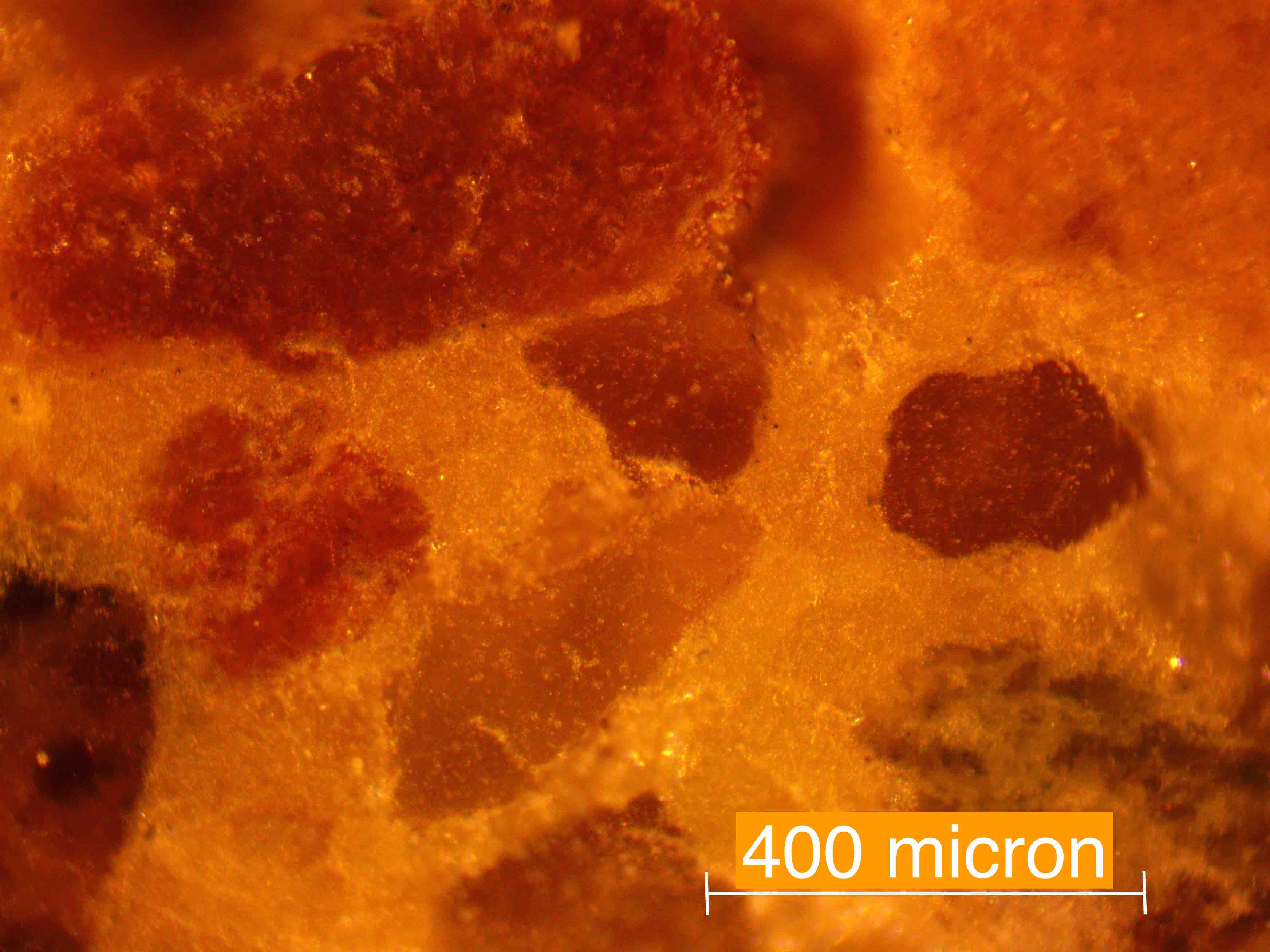}
\caption{Microscopy study of sulfur concrete on 400 $\mu$m scale with compositions of (a) 50\% sulfur and 50\% Martian soil simulant (b) 25\% sulfur and 75\% regular sand and a maximum particle size of 1~mm}
\label{m2}
\efi

\subsection{Three-point-bending Fracture Test}

To complete the mechanical characterization of MC, its fracturing behavior is studied in this section and the next. Beam specimens with nominal dimensions 25.4$\times$25.4$\times$127~mm (1$\times$1$\times$5~in) were cast to perform three-point-bending (TPB) tests. The beam specimens featured a half-depth notch at midspan cut with a diamond coated band-saw machine. Testing notched samples is customary in fracture mechanics to control the fracture onset and to capture post-peak behavior. Dimension and weight measurements were recorded on specifically optimized TPB protocols. Centerline on top of specimen, and support lines at the bottom were pre-marked then aligned within the servo-hydraulic load frame, which had a capacity of 22.2 kN (5 kip). The adopted TPB test setup is shown in Fig.~\ref{tpb}a. The nominal span (distance between bottom supports) was 101.6~mm (4~in). An extensometer sensor was glued to the bottom of the specimens with the notch in between its two feet. After applying a pre-load of up to 5$\%$ of the expected peak, the specimens were loaded in crack mouth opening displacement (CMOD) control with a loading rate of 0.0001~mm/sec, which was increased in the post-peak section to limit the total testing time while ensuring a fully recorded softening behavior. Typical crack propagation and fracture surface after failure are presented in Fig.~\ref{tpb}b\&c. The crack starts at the notch tip and develops upward along the ligament. 


Notched (50\%) fracture test stress-strain curves of MC with sulfur ratio in the range of 40\% $\sim$ 60\% are plotted in Fig.~\ref{alle}b. The nominal tensile flexural stress is calculated as $\sigma = 3PL/2bh^2$, where $P$ is load, and $L$, $b$, and $h$ are span, width, and depth of the specimen respectively; the nominal strain is calculated as $\epsilon$=CMOD/$h$. The optimal percentage of sulfur is found to be 50$\%$ ($\pm$ 2.5\%) which gives a nominal flexural strength of approximately 1.65~MPa, and it agrees with the optimal percentage determined from unconfined compression tests. The highest nominal flexural strength obtained is 2.3~MPa reached by one of the two recast 50\% sulfur batches, as shown in Fig.~\ref{pertpb}a. It must be observed that nominal flexural strength and flexural nominal stress-strain curves are not material properties, due to the presence of the notch, and they are calculated here only for comparison purposes. The typical material property that can be calculated from TPB test is the fracture energy, defined as the energy per unit area needed to create a unit stress-free fracture area. By adopting the work-of-fracture method \cite{Bazant1996} the fracture energy is computed by dividing the area under the load vs. stroke curve by the ligament area. The highest average total fracture energy is as well reached by the recast Martian Concrete with 50\% sulfur with a value of 67 J/m$^2$, as shown in Fig.~\ref{pertpb}b. When mixed with lower or higher sulfur ratio than 50\%, MC has lower fracture energies, see Fig.~\ref{alle}b and Fig.~\ref{pertpb}b. Same as for compressive strength, recast and applying pressure can as well improve material flexural strength thanks to more compact sulfur bonds. 

\bfi
\centering
(a) \includegraphics[height=2.5in,valign=t]{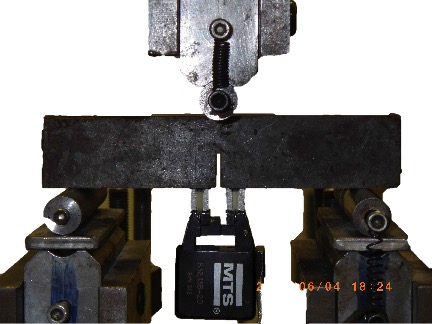}
(b)  \includegraphics[height=2.1in,valign=t]{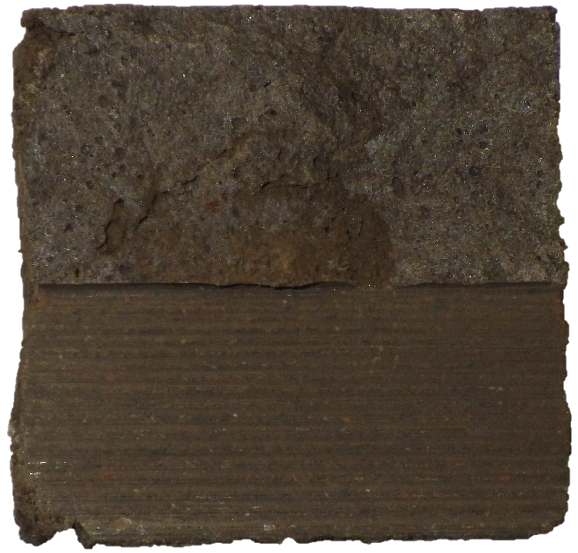}\\
(c)  \includegraphics[height=1.25in,valign=t]{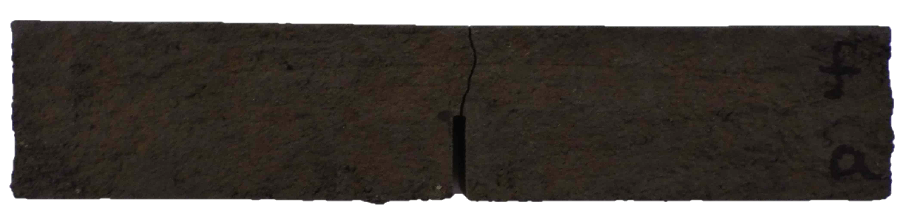}
\caption{(a) Three point bending (TPB) test setup, (b) fracture surface and (c) typical crack propagation after bending test of Martian Concrete}
\label{tpb}
\efi

\bfi
\centering
(a) \includegraphics[width=3.25in,valign=t]{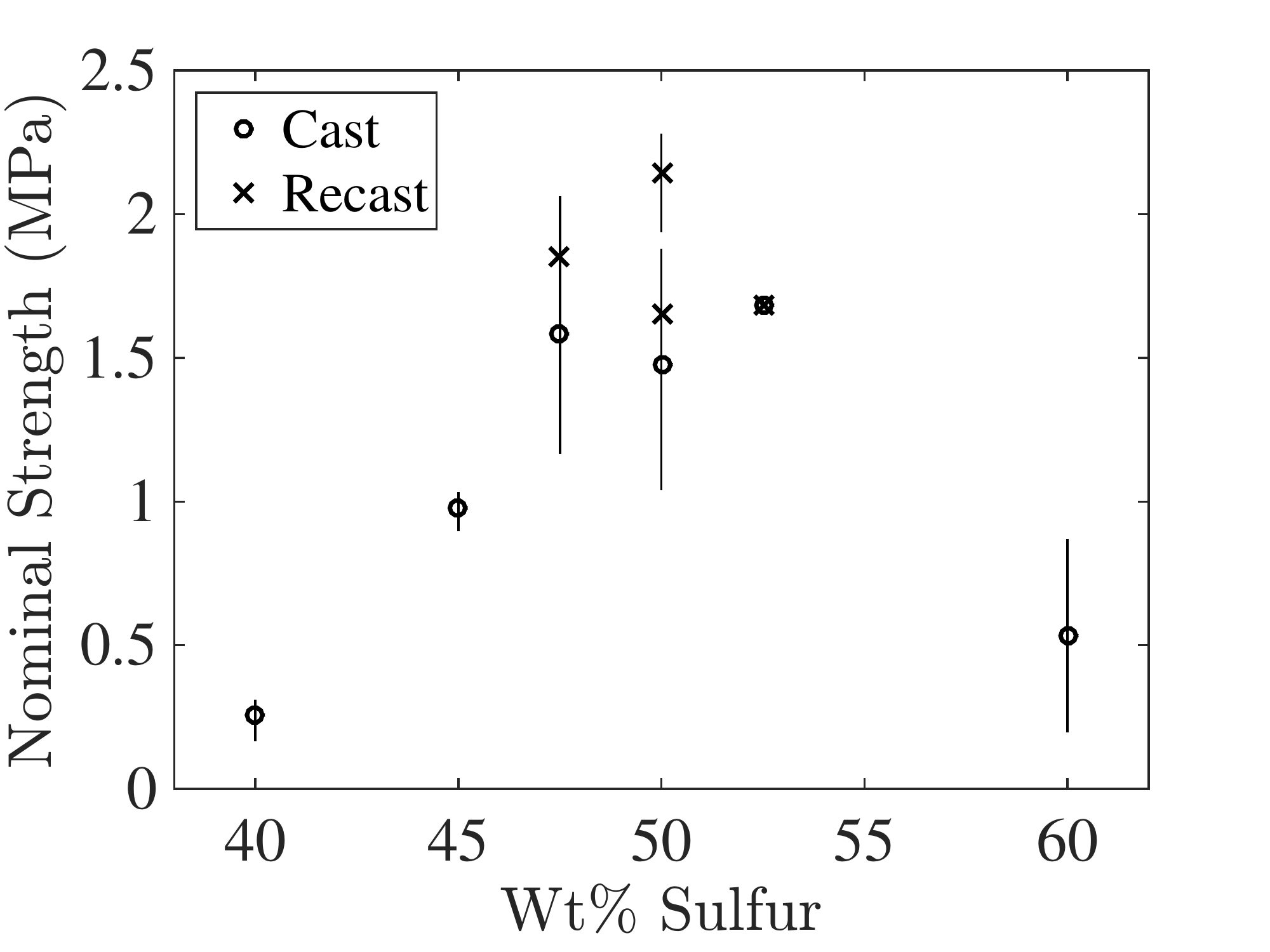}
(b)  \includegraphics[width=3.25in,valign=t]{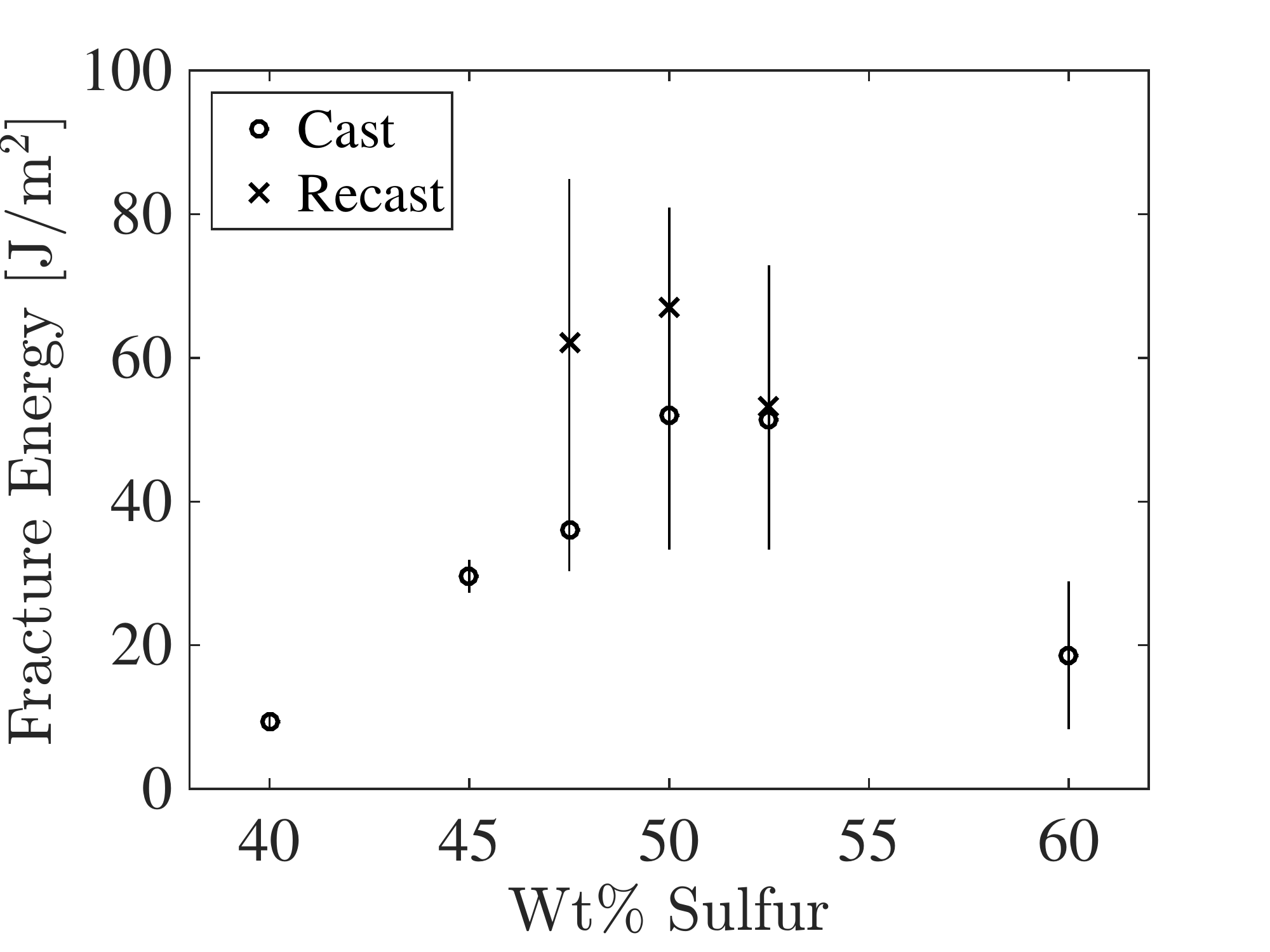}
\caption{Best percentage of sulfur for Martian Concrete by TPB test results (a) nominal flexural strength, and (b) fracture energy}
\label{pertpb}
\efi

\subsection{Splitting and Modulus of Rupture Tests}

Splitting tests on 25.4~mm (1~in) cubes were performed by the same load frame as for compression. Roughly 1~mm diameter bars were placed on the top and at the bottom of the specimen. A loading rate of 0.003~mm/s was applied until failure of the specimen at peak load. Only recast Martian Concrete with 47.5$\%$, 50$\%$, and 52.5$\%$ were tested, and provided splitting tensile strength of 3.6~MPa $\pm$ 30\%, 3.9~MPa $\pm$ 28\%, and 2.72~MPa $\pm$ 26\% respectively. The splitting tensile strength is calculated as $\sigma = 2P/\pi bh$, where $P$ is load, $b$ and $h$ are the depth and height of the cube specimen respectively. In agreement with compression and TPB test results, splitting tests again confirm that MC with 50\% of sulfur have the highest performance. The splitting nominal stress-strain curves, until failure at peak load, of the optimum MC are shown in Fig.~\ref{val}b, where nominal strain is calculated as vertical displacement divided by the specimen height.

Modulus of rupture (MOR) tests were carried out for MC with the optimum mix, 50\% sulfur and 50\% Martian soil simulant. Unnotched beams with dimensions 25.4$\times$25.4$\times$127~mm (1$\times$1$\times$5~in) were tested for MOR using the aforementioned machine and setup for notched TPB but by stroke control with loading rate 0.001 mm/s. The developed MC has an average MOR value of 7.24 MPa, see Fig.~\ref{val}a. The nominal MOR stress is calculated as $\sigma = 3PL/2bh^2$, where $P$ is load, $L$, $b$, and $h$ are span, width, and depth of the specimen respectively; the nominal strain is calculated as vertical displacement divide by specimen depth.

\section{Lattice discrete Particle Model Simulations}

For design and analysis purposes it is important to formulate and validate a computational model for the simulation of Martian Concrete. This is pursued within the theoretical framework of the Lattice Discrete Particle Model (LDPM).

\begin{figure}[h]
  \centering 
\includegraphics[width=2.5 in]{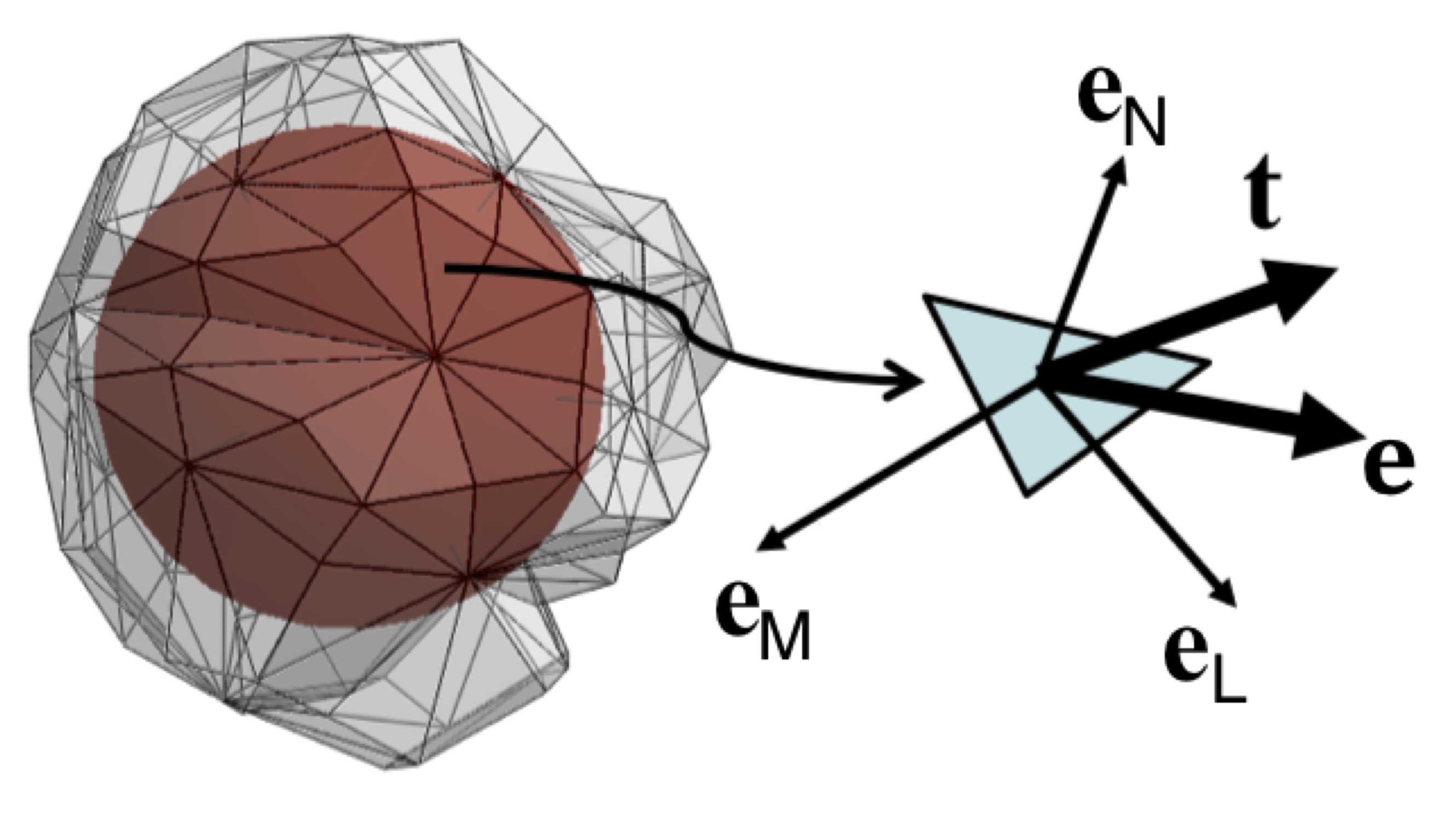}
\caption{One LDPM Cell around an aggregate piece. }
    \label{ldpm}
\end{figure}

In 2011, building on previous work \cite{CuBaCe2003a, CuBaCe2003b, CuBaCe2006}, Cusatis and coworkers \cite{Cusatis2011a,Cusatis2011b} developed LDPM, a mesoscale discrete model that simulates the mechanical interaction of coarse aggregate pieces embedded in a binding matrix. The geometrical representation of concrete mesostructure is constructed by randomly introducing and distributing spherical shaped coarse aggregate pieces inside the volume of interest and zero-radius aggregate pieces on the surface. Based on the Delaunay tetrahedralization of the generated particle centers, a three-dimensional domain tessellation creates a system of polyhedral cells (see Fig.~\ref{ldpm}) interacting through triangular facets and a lattice system. The full description of LDPM geometry is reported in Cusatis. et. al. \cite{Cusatis2011a,Cusatis2011b}.

In LDPM, rigid body kinematics is used to describe the deformation of the lattice particle system and the displacement jump, $\llbracket \mathbf{u}_{C} \rrbracket$, at the centroid of each facet is used to define measures of strain as 
\begin{equation}
\label{eps} e_{N}=\frac{\mathbf{n}^\mathrm{T} \llbracket \mathbf{u}_{C} \rrbracket}{\ell};
\;\;\; e_{L}=\frac{\mathbf{l}^\mathrm{T} \llbracket \mathbf{u}_{C} \rrbracket} {\ell};
\;\;\; e_{M}=\frac{ \mathbf{m}^\mathrm{T} \llbracket \mathbf{u}_{C} \rrbracket}{\ell}
\end{equation}
where $\ell=$ interparticle distance; and $\mathbf{n}$, $\mathbf{l}$, and
$\mathbf{m}$, are unit vectors defining a local system of reference attached to each facet. A vectorial constitutive law governing the material behavior is imposed at the centroid of each facet. In the elastic regime, the normal and shear stresses are proportional to the corresponding strains: $t_{N}= E_N e^*_{N} =E_N (e_{N}-e^0_{N});~ t_{M}= E_T e^*_{M} = E_T (e_{M}-e^0_{M});~ t_{L}= E_T e^*_{L} = E_T (e_{L}-e^0_{L})$, where $E_N=E_0$, $E_T=\alpha E_0$, $E_0=$ effective normal modulus, and $\alpha=$ shear-normal coupling parameter; and $e^0_{N}$, $e^0_{M}$, $e^0_{L}$ are mesoscale eigenstrains (if any present). For stresses and strains beyond the elastic limit, the LDPM formulation considers the following nonlinear mesoscale phenomena \cite{CuBaCe2003a, CuBaCe2003b, Cusatis2011a}: (1) fracture and cohesion; (2) compaction and pore collapse; and (3) internal friction.
 
\textbf{Fracture and cohesion due to tension and tension-shear.} For tensile loading ($e^*_N>0$), the fracturing behavior is formulated through effective strain, $e^* = \sqrt{e_N^{*2}+\alpha (e_M^{*2} + e_L^{*2})}$, and stress, $t = \sqrt{{ t _{N}^2+  (t _{M}^2+t _{L}^2) / \alpha}}$, which define the normal and shear stresses as \mbox{$t _{N}= e_N^*(t / e^*)$}; \mbox{$t _{M}=\alpha e^*_{M}(t / e^*)$}; \mbox{$t _{L}=\alpha e^*_{L}(t / e^*)$}. The effective stress $t$ is incrementally elastic ($\dot{t}=E_0\dot{e}$) and must satisfy the inequality $0\leq t \leq \sigma _{bt} (e, \omega) $ where $\sigma_{bt} = \sigma_0(\omega) \exp \left[-H_0(\omega)  \langle e-e_0(\omega) \rangle / \sigma_0(\omega)\right]$, $\langle x \rangle=\max \{x,0\}$, and $\tan(\omega) =e^* _N / \sqrt{\alpha} e^* _{T}$ = $t_N \sqrt{\alpha} / t_{T}$, and $e_T^*=\sqrt{e_M^{*2} + e_L^{*2}}$. The post peak softening modulus is defined as $H_{0}(\omega)=H_{t}(2\omega/\pi)^{n_{t}}$, where $H_{t}$ is the softening modulus in pure tension ($\omega=\pi/2$). LDPM provides a smooth transition between pure tension and pure shear ($\omega=0$) with parabolic variation for strength given by $\sigma_{0}(\omega )=\sigma _{t}r_{st}^2\Big(-\sin(\omega) + \sqrt{\sin^2(\omega)+4 \alpha \cos^2(\omega) / r_{st}^2}\Big) / [2 \alpha \cos^2(\omega)]$, where $r_{st} = \sigma_s/\sigma_t$ is the ratio of shear strength to tensile strength. 

\textbf{Compaction and pore collapse from compression.} Normal stresses for compressive loading ($e^*_N<0$) must satisfy the inequality $-\sigma_{bc}(e_D, e_V)\leq t_N \leq 0$, where $\sigma_{bc}$ is a strain-dependent boundary depending on the volumetric strain, $e_V$, and the deviatoric strain, $e_D=e_N-e_V$. The volumetric strain is computed by the volume variation of the Delaunay tetrahedra as $e_V= \Delta V/ 3V_0$ and is assumed to be the same for all facets belonging to a given tetrahedron. Beyond the elastic limit, $-\sigma_{bc}$ models pore collapse as a linear evolution of stress for increasing volumetric strain with stiffness $H_{c}$ for $-e_V \leq e_{c1} = \kappa_{c0} e_{c0}$: $\sigma_{bc} = \sigma_{c0} + \langle-e_V-e_{c0}\rangle H_c(r_{DV})$; $H_c(r_{DV})=H_{c0}/(1 + \kappa_{c2} \left\langle r_{DV} - \kappa_{c1} \right\rangle)$; $\sigma_{c0}$ is the mesoscale compressive yield stress; $r_{DV}=e_D/e_V$ and $\kappa_{c1}$, $\kappa_{c2}$ are material parameters. Compaction and rehardening occur beyond pore collapse ($-e_V \geq e_{c1}$). In this case one has $\sigma_{bc} = \sigma_{c1}(r_{DV})$ $\exp \left[( -e_{V}-e_{c1} ) H_c(r_{DV})/\sigma_{c1}(r_{DV}) \right]$ and $\sigma_{c1}(r_{DV}) = \sigma_{c0} + (e_{c1}-e_{c0}) H_c(r_{DV})$. 

\textbf{Friction due to compression-shear.} For compression dominated loading conditions ($e^*_N<0$), the incremental shear stresses are computed as  $\dot{t}_M=E_T(\dot{e}^*_M-\dot{e}^{*p}_M)$ and \mbox{$\dot{t}_L=E_T(\dot{e}^*_L-\dot{e}^{*p}_L)$}, where  \mbox{$\dot{e}_M^{*p}=\dot{\xi} \partial \varphi / \partial t_M$}, \mbox{$\dot{e}_L^{*p}=\dot{\xi} \partial \varphi / \partial t_L$}, and $\xi$ is the plastic multiplier with loading-unloading conditions  $\varphi \dot{\xi} \leq 0$ and $\dot{\xi} \geq 0$. The plastic potential is defined as \mbox{$\varphi=\sqrt{t_M^2+t_L^2} - \sigma_{bs}(t_N)$}, where the nonlinear frictional law for the shear strength is assumed to be $\sigma_{bs} = \sigma_s + (\mu_0 - \mu_\infty)\sigma_{N0}[1 - \exp(t_N / \sigma_{N0})] - \mu_\infty t_N$; $\sigma_{N0}$ is the transitional normal stress; $\mu_0$ and $\mu_\infty$ are the initial and final internal friction coefficients.  

Each meso-level parameter in LDPM governs part of the mechanical material behavior. The normal elastic modulus, which refers to the stiffness for the normal facet behavior, $E_0$, , along with the coupling parameter $\alpha$, govern LDPM response in the elastic regime. Approximately, the macro scale Young's modulus $E$ and Poisson's ratios $\nu$ can be calculated as $E = E_0(2+3\alpha)/(4+\alpha)$ and $\nu = (1-\alpha)/(4+\alpha)$. Typical concrete Poisson's ratio of about 0.18 is obtained by setting $\alpha$ = 0.25 \cite{Cusatis2011b}. The tensile strength, $\sigma_t$, and characteristic length, $\ell_{t}$, govern the strain softening behavior due to fracture in tension of LDPM facets \cite{Cusatis2011b}, with the relation $G_t = \ell_{t} \sigma_t^2 /2E_0$, where $G_t$ is the mesoscale fracture energy. Calibration of $\sigma_t$ and $\ell_t$ is typically achieved by fitting experimental data, e.g. the nominal stress-strain curves of TPB tests. The yielding compressive stress, $\sigma_{c0}$, defines the behavior of the facet normal component under compression. The softenig exponent, $n_t$, governs the interaction between shear and tensile behavior during softening at the facet level and it governs the macroscopic compressive behavior at high confinement. One obtains more ductile behavior in both compression and tension by increasing $n_t$, however the increase is more pronounced in compression than in tension. The initial internal friction, $\mu_0$, mainly govern the mechanical response in compression at low confinement and have no influence on tensile behavior. Descriptions of effects and functions of other LDPM mesoscale parameters and further discussions can be found in Cusatis et. al. \cite{Cusatis2011b} and Wan et. al. \cite{WanUHPCI}.


LDPM has been utilized successfully to simulate cementitious concrete behavior under various loading conditions \cite{Cusatis2011a,Cusatis2011b}. Furthermore, the framework has been extended to properly account for fiber reinforcement \cite{Schauffert2012I,Schauffert2012II} and has the ability to simulate the mechanical behavior of ultra high performance concrete (UHPC) \cite{Smith2014, WanConcreep2015, WanUHPCI} and long term behavior of concrete with fastening applications \cite{Concreep2}. 

Although Martian Concrete has sulfur bonds instead of calcium-silicate-hydrate gels, it shares with cementitious concrete the heterogeneous internal structure, which is the basis of the LDPM formulation. Thus, LDPM is adopted to simulate the mechanical behavior of the Martian Concrete. The numerical simulations presented in this paper were performed with the software MARS, a multi-purpose computational code, which implements LDPM, for the explicit dynamic simulation of structural performance \cite{MARS2009}. As aforementioned, the particle size of aggregate in MC is shifted to lower ends after casting, however, the exact distribution cannot be obtained and simulating the smallest particles would result in significantly high computation cost. Thus, the discrete particles are generated randomly with aggregate piece of 0.5~to~1~mm and Fuller's law to the power 1/2 for each type of specimen. The utilized mesoscale parameters for MC with the best sulfur ratio (50\%) are listed in Table 2. The TPB experimental data was primarily utilized to calibrate the LDPM parameters governing elastic as well as fracture behavior, which include normal modulus, tensile strength, shear strength ratio, tensile characteristic length, and softening exponent. Note that the normal modulus is calibrated by the TPB test data because the nominal strain (CMOD/$h$) is directly measured on the specimen, while all other tests include the effect of the test machine compliance. Compression experimental data was then used to calibrate the shear strength, the softening exponent and the initial internal friction. The other parameters' values, relevant to confined compressive behavior, are determined based on calibrated sets for typical concrete materials available in the literature \cite{Cusatis2011b, WanUHPCI} and are assumed to work also for Martian Concrete in absence of specific experimental data. The adopted values are densification ratio = 1, asymptotic friction = 0, transitional stress = 300~MPa, volumetric deviatoric coupling coefficient = 0, deviatoric strain threshold ratio = 1, and deviatoric damage parameter = 5. After all LDPM parameters had been calibrated and determined, prediction simulations for unnotched TPB tests and splitting tests were carried out and compared to experimental data as validation. 

The LDPM simulation setup, typical failure type and crack propagation of notched TPB, unconfined compression, splitting, and unnotched TPB tests are shown in Fig,~\ref{ldpmtpb},~\ref{ldpmcube},~\&~\ref{ldpmbdun} respectively. Note that in the notched and unnotched TPB simulation setup (Fig.~\ref{ldpmtpb} \& \ref{ldpmbdun}), the specimen is composed of lattice discrete particles at the center and classical elastic finite elements on the two sides, where only elastic deformation is expected to occur, to save computational time. In the unconfined compression test simulation, high friction parameters for typical concrete-steel slippage interaction \cite{Cusatis2011b} are utilized: $\mu_s$ = 0.13, $\mu_d$ = 0.015, and $s_0$ = 1.3~mm, to simulate friction between the specimen ends and the steel loading platens, assuming a slippage-dependent friction coefficient formulated as $\mu(s) = \mu_d + (\mu_s-\mu_d)s_0/(s_0+s)$. The fitted stress-strain curves can be found in Fig.~\ref{cal}~\&~\ref{val}. Fig.~\ref{cal}a shows the nominal stress-strain curves for 50\% notched TPB tests and the material has total fracture energy, $G_F$, of 67.0 J/m$^2$. The mesoscale initial fracture energy calculated from LDPM parameters, $G_t = \ell_{t} \sigma_t^2 /2E_0$ = 37.6 J/m$^2$, is approximately half of $G_F$. This is due to the fact that even under macroscopic mode I fracture the mesoscale response is characterized by both shear and tension. Fig.~\ref{cal}b presents the experimental and simulated stress-strain curves of unconfined compression test. Young's modulus $E$ is back calculated as the aforementioned equation $E = E_0(2+3\alpha)/(4+\alpha)$ and has an average value of 6.5 GPa. This value is then used to remove the machine compliance in experimental compression test data. 

Brittle failure is observed both in experiments and simulations for unnotched TPB and splitting tests, as shown in Fig.~\ref{val}~a~\&~b respectively. The compliance in splitting and unnotched TPB experimental data is removed according to calibrated simulations. As pure predictions, the simulation peaks highly agree with the average strengths of the experiments. This indicates the superior ability of LDPM to simulate and predict the mechanical behavior of not only cement based concrete but also the novel waterless Martian concrete materials. 

\begin{table}[ht]
\centering
\begin{tabular}{l c}
\multicolumn{2}{l}{Table 2: Parameters for Martian Concrete LDPM Simulations}\\
\hline
NormalModulus [GPa]  & 10 \\
DensificationRatio [-]  & 1 \\
TensileStrength [-] [MPa] & 3.7\\
YieldingCompressiveStress [MPa] & 300 \\
ShearStrengthRatio [-]  & 4 \\
TensileCharacteristicLength [mm] & 55 \\
SofteningExponent [-]  & 0.2 \\
InitialHardeningModulusRatio [-]  &0.12 \\
TransitionalStrainRatio [-]  &4 \\
InitialFriction [-]  &0.1 \\
\hline
\end{tabular}
\end{table}

\bfi
\centering
\includegraphics[height=2.5in,valign=t]{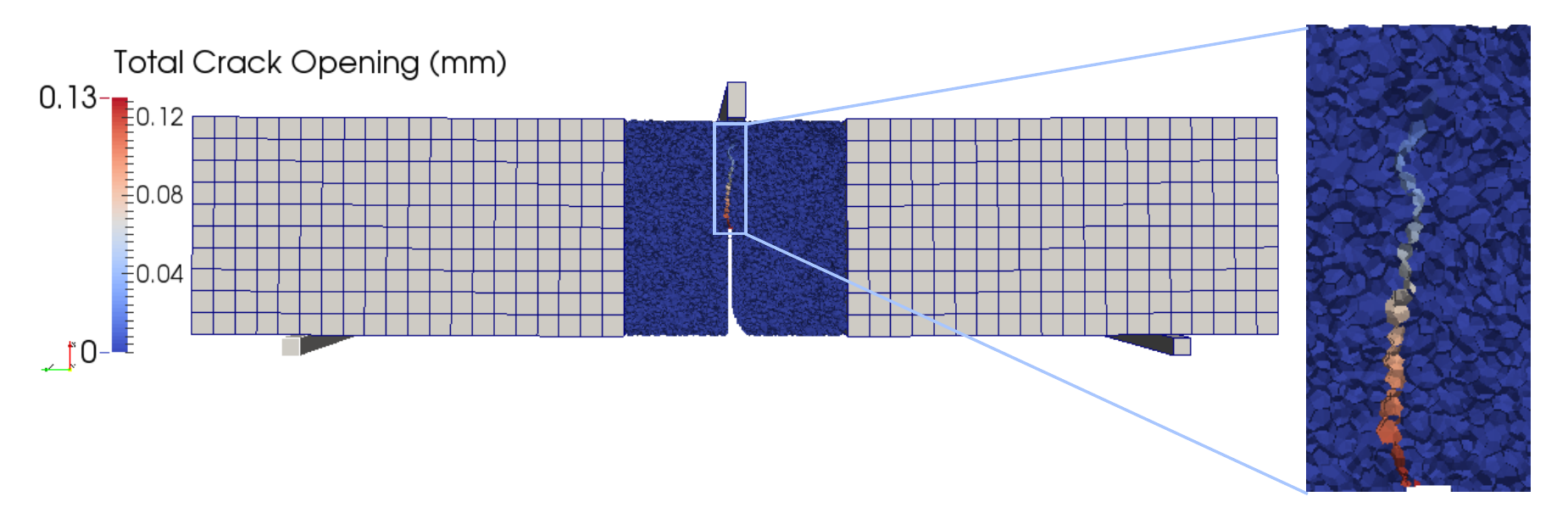}
\caption{LDPM simulation of notched TPB test setup and zoomed-in view of crack propagation}
\label{ldpmtpb}
\efi

\bfi
\centering
(a) \includegraphics[height=2.7in,valign=t]{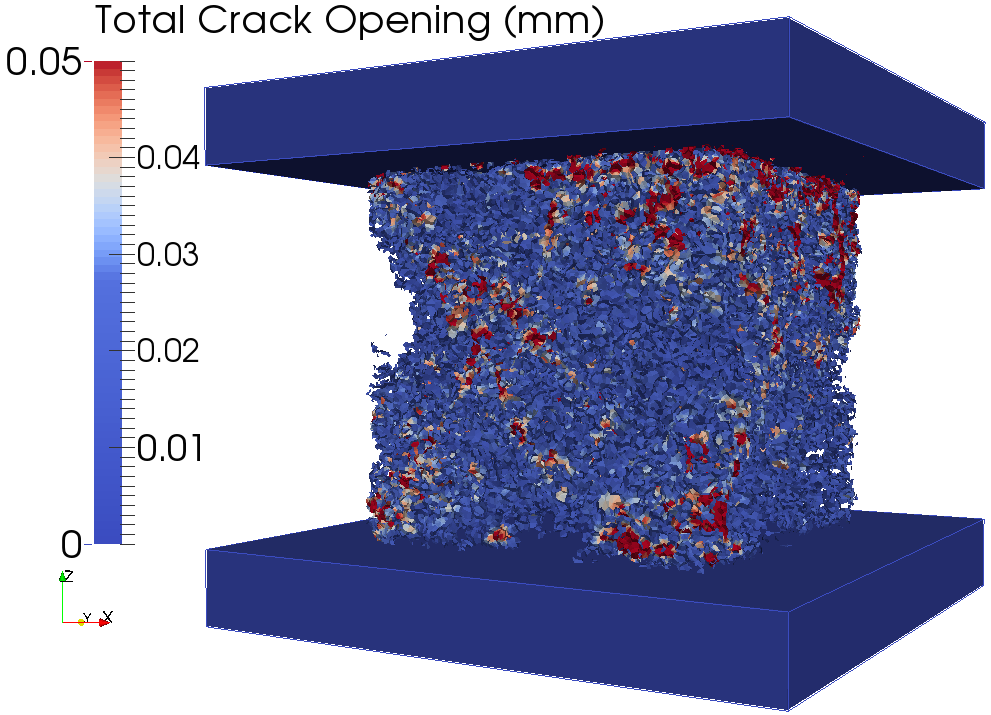}
(b) \includegraphics[height=2.8in,valign=t]{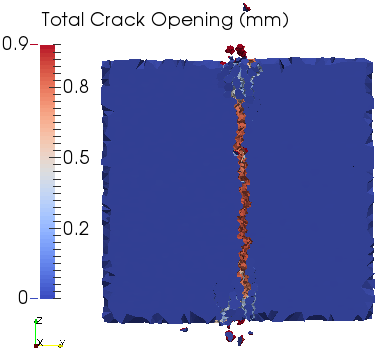}
\caption{LDPM simulation of typical crack propagation in (a) unconfined compression test and (b) splitting (Brazilian) test}
\label{ldpmcube}
\efi

\bfi
\centering
\includegraphics[height=2in,valign=t]{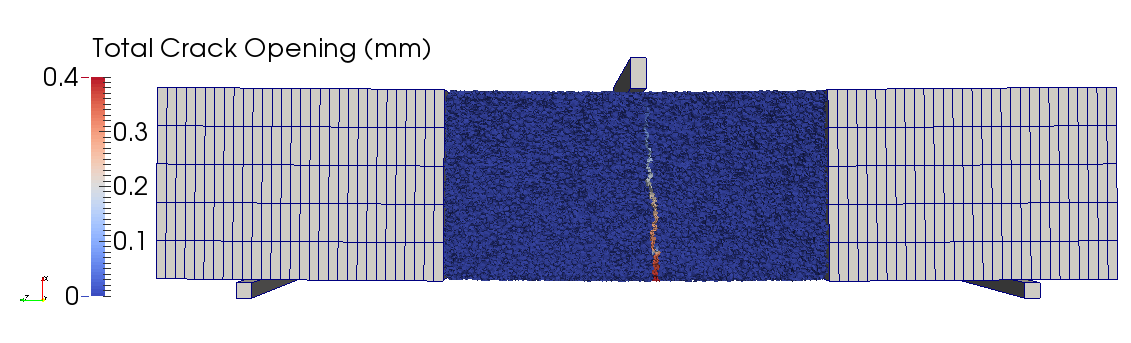}
\caption{LDPM simulation of unnotched TPB test setup and typical crack propagation}
\label{ldpmbdun}
\efi

\bfi
\centering
(a) \includegraphics[width=3.25in,valign=t]{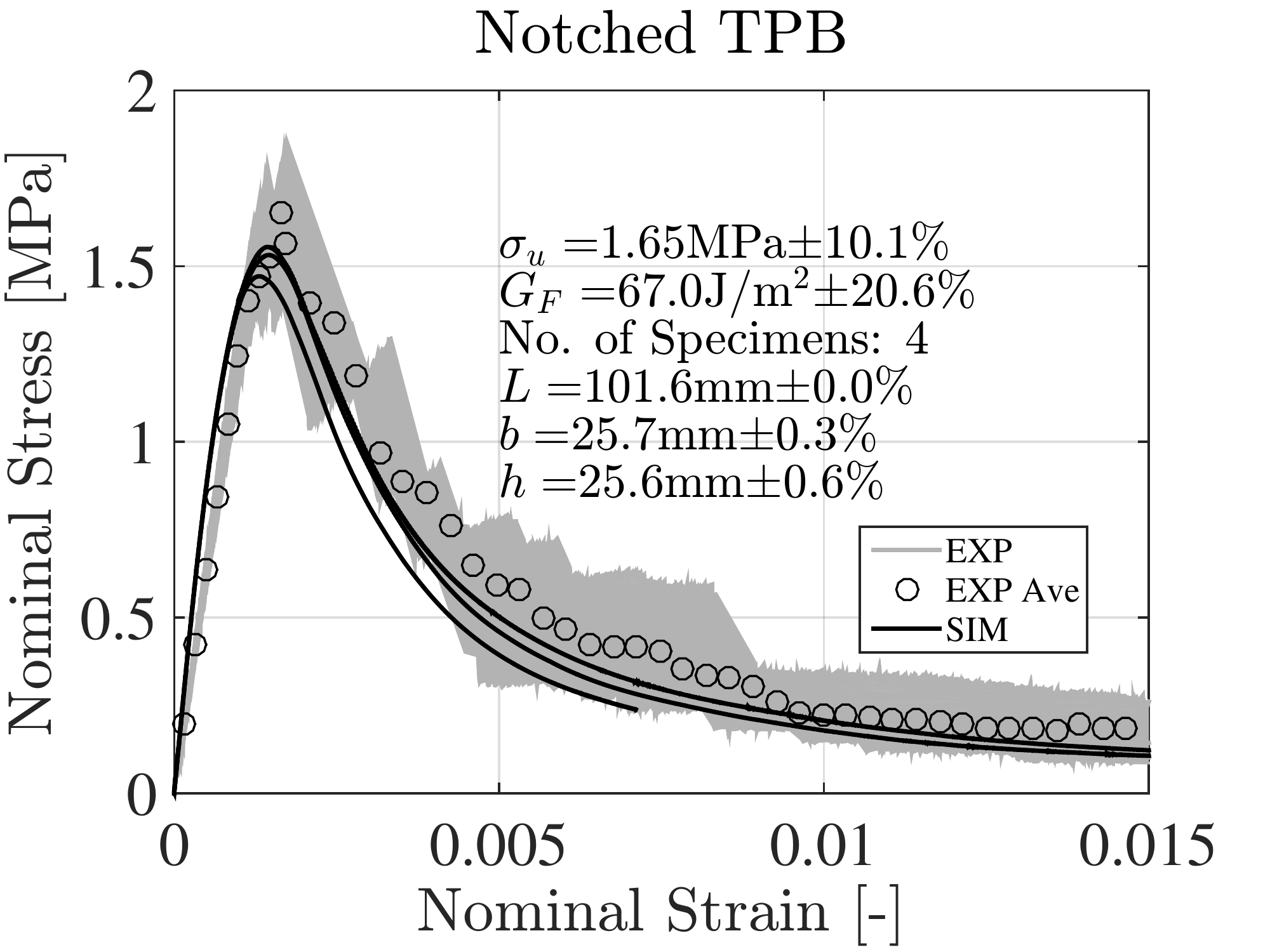}
(b) \includegraphics[width=3.25in,valign=t]{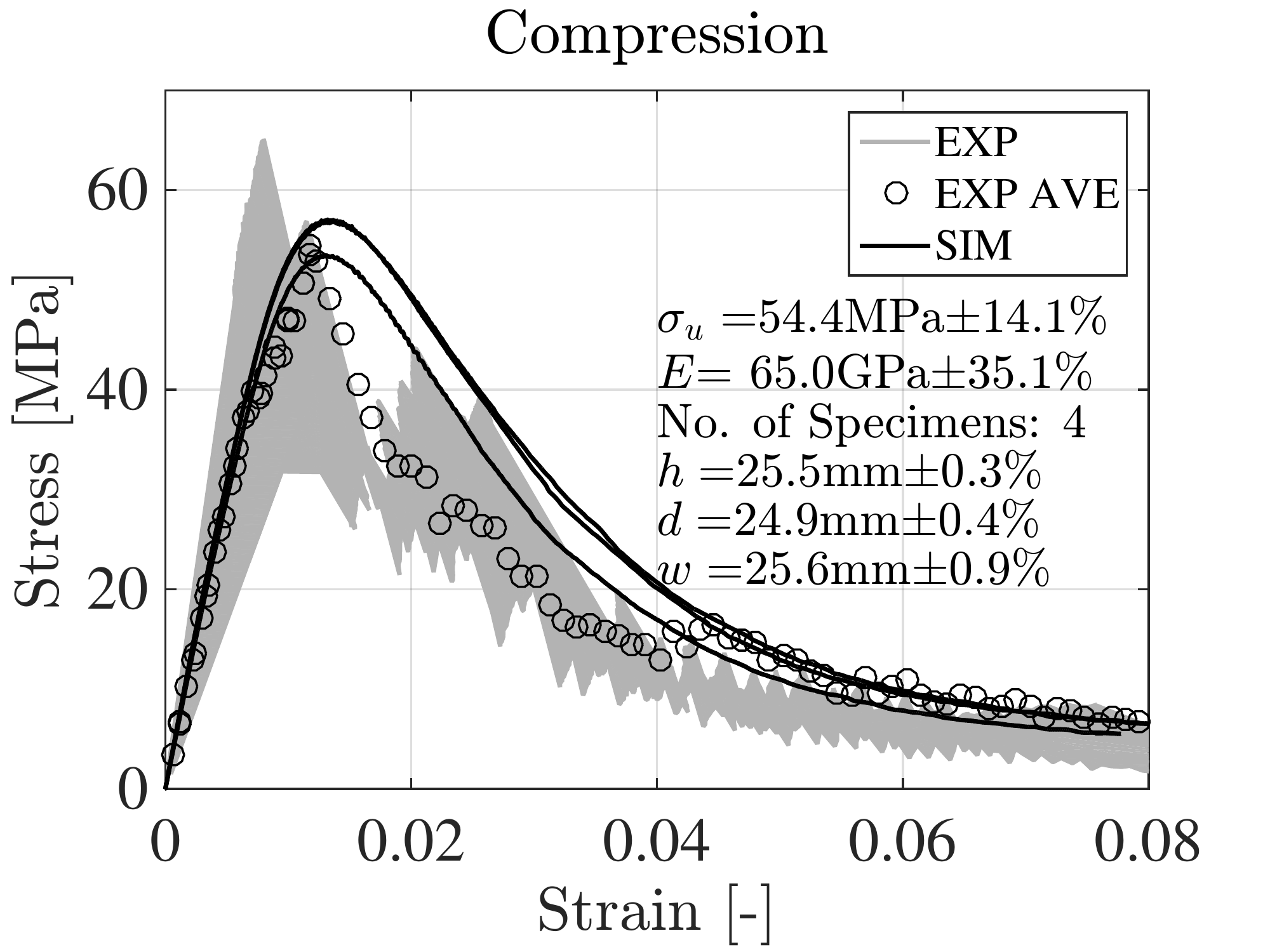}
\caption{Experimental results and LDPM simulations for calibration and validation: (a) 50\% notched three-point-bending tests (b) unconfined compression tests}
\label{cal}
\efi

\bfi
\centering
(a) \includegraphics[width=3.25in,valign=t]{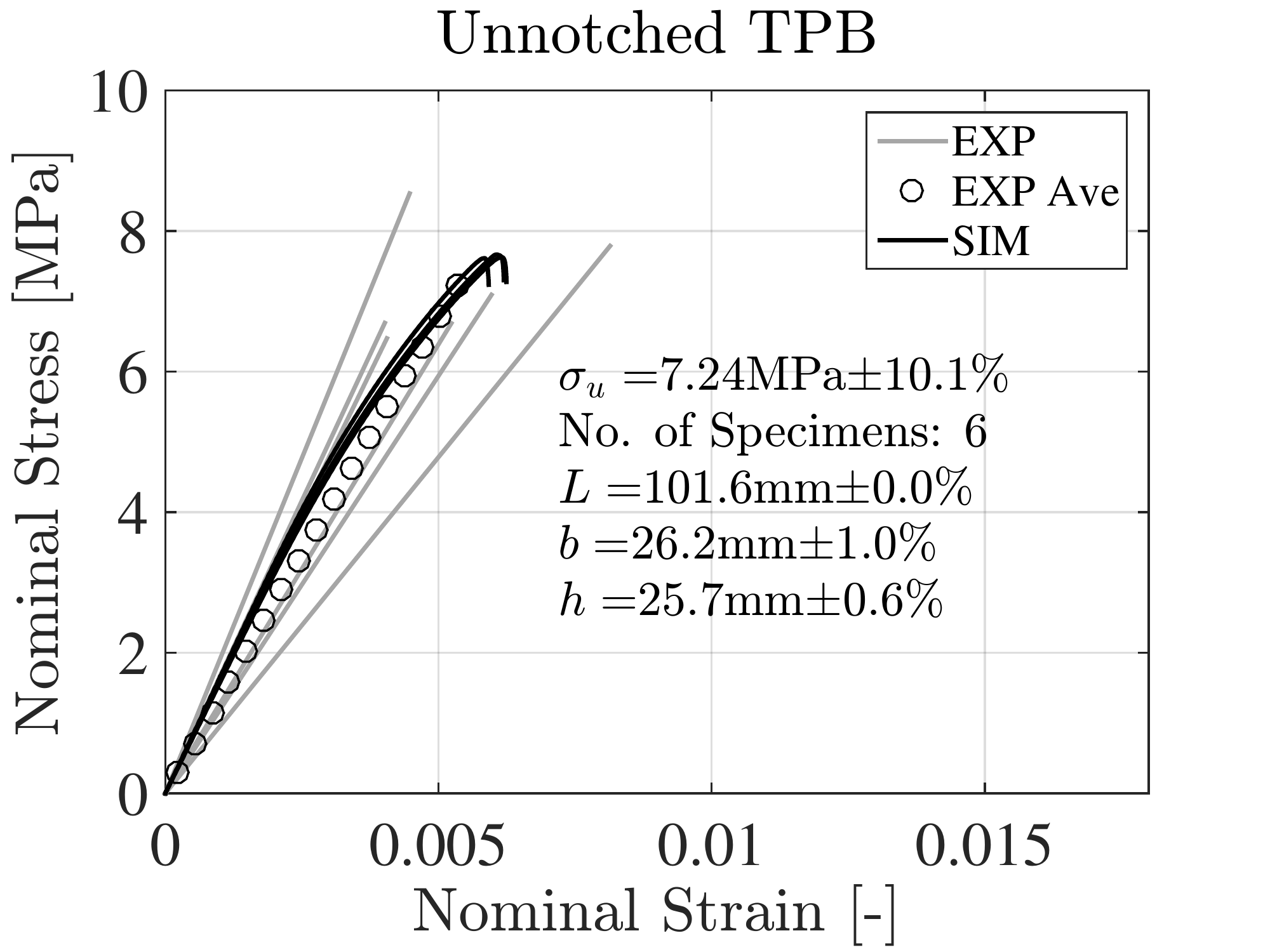}
(b) \includegraphics[width=3.25in,valign=t]{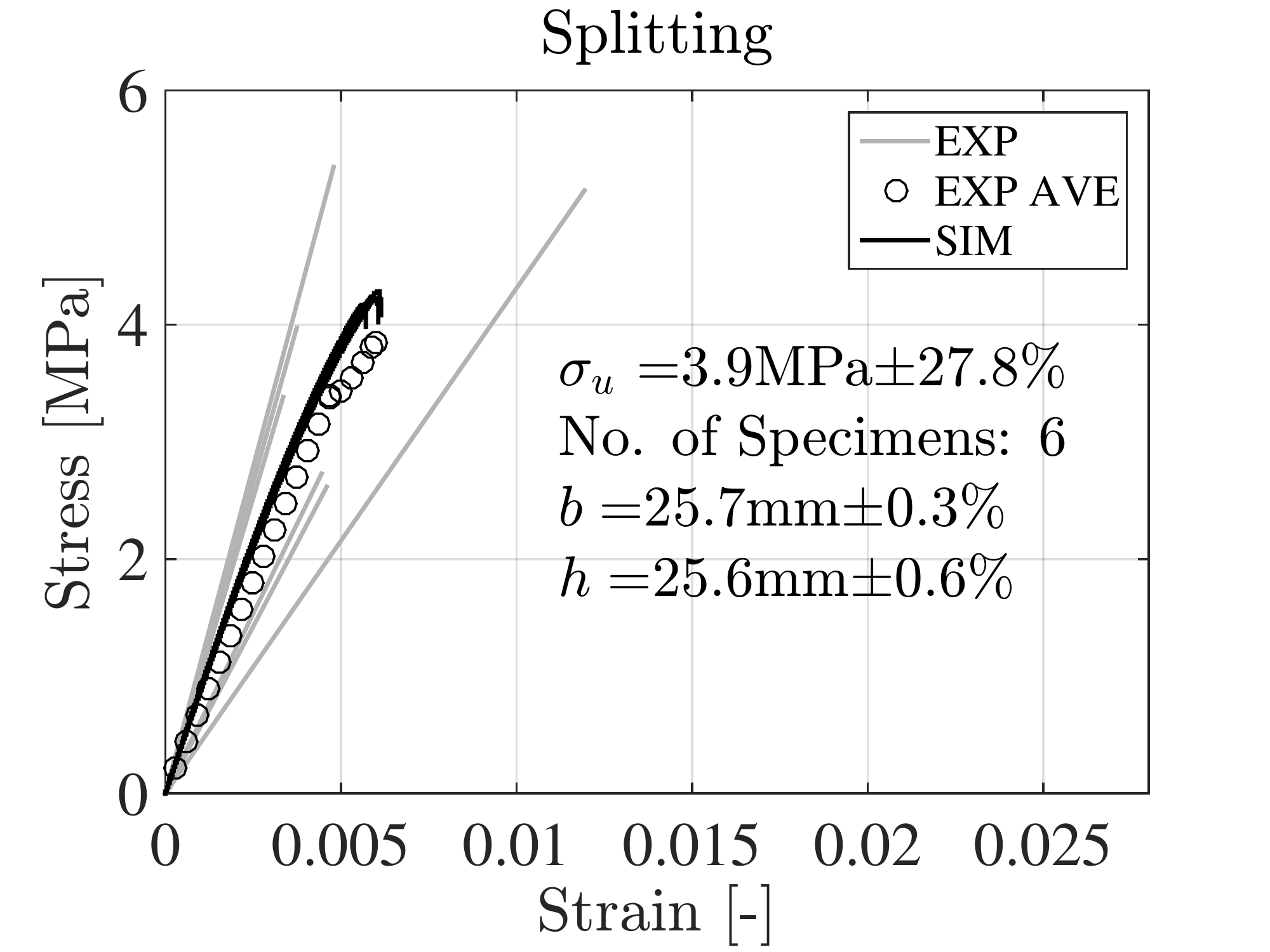}
\caption{Experimental results and LDPM simulations for validation: (a) unnotched three-point-bending tests (b) splitting tests}
\label{val}
\efi

\section{Summary and Conclusions}

In conclusion, the developed sulfur based Martian Concrete is feasible for construction on Mars for its easy handling, fast curing, high strength, recyclability, and adaptability in dry and cold environments. Sulfur is abundant on Martian surface and Martian regolith simulant is found to have well graded particle size distribution to ensure high strength mix. Both the atmospheric pressure and temperature range on Mars are adequate for hosting sulfur concrete structures. Based upon the experimental and numerical results presented in this paper, the following conclusions can be drawn:

\bi
\item The best mix for producing Martian Concrete (MC) is 50\% sulfur and 50\% Martian soil simulant with maximum aggregate size of 1~mm. The developed MC can reach compressive strength higher than 50~MPa. 
\item The optimum particle size distribution (PSD) of Martian regolith simulant is found to play a role in achieving high strength MC compared to sulfur concrete with regular sand.
\item The rich metal elements in Martian soil simulant are found to be reactive with sulfur during hot mixing, possibly forming sulfates and polysulfates, which further increases MC strength. Simultaneously, the particle size distribution of aggregate is shifted to lower ends, resulting in less voids and higher performance of the final mix. 
\item With the advantage of recyclability, recast of MC can further increase the material's overall performance. 
\item Applying pressure during casting can also increase the final strength of MC. Sulfur shrinks when is cooled down. By reducing the mixture's volume during casting, the number and size of cavities of the final product are decreased.
\item Although developed for conventional cementitious concrete, the Lattice Discrete Particle Model (LDPM) shows also excellent ability in simulating the mechanical behavior of MC under various loading conditions.
\ei

\section*{Acknowledgement}
The work was financially supported with Northwestern University internal funding. The authors would like to thank laboratory coordinator Dave Ventre and undergraduate student Timothy Clark for their contribution to material preparation in the experimental campaign.

\end{document}